\documentclass[reprint,twocolumn,superscriptaddress,preprintnumbers,amsmath,amssymb,aps,prd,floatfix]{revtex4-2}

\usepackage[utf8]{inputenc}

\usepackage{mathtools}
\usepackage{amsfonts}
\usepackage{mathrsfs}
\usepackage{bm}
\usepackage[normalem]{ulem}
\usepackage{bbold}
\usepackage{braket}
\usepackage{slashed}
\usepackage{dsfont}
\usepackage{float}
\usepackage{dcolumn}
\usepackage{amssymb,amsmath}

\usepackage{graphicx}
\usepackage{color}
\usepackage{array}

\usepackage{placeins}
\usepackage{booktabs}

\usepackage{xspace}
\usepackage{hyperref}
\usepackage[nameinlink]{cleveref}
\usepackage{bookmark}
\usepackage{units}

\setkeys{Gin}{width=0.48\textwidth}

\usepackage{booktabs}
\usepackage{multirow}

\newcolumntype{C}{>{$}c<{$}}
\AtBeginDocument{
	\heavyrulewidth=.08em
	\lightrulewidth=.05em
	\cmidrulewidth=.03em
	\belowrulesep=.65ex
	\belowbottomsep=0pt
	\aboverulesep=.4ex
	\abovetopsep=0pt
	\cmidrulesep=\doublerulesep
	\cmidrulekern=.5em
	\defaultaddspace=.5em
}

\graphicspath{{./figures/}}

\usepackage{xifthen}
\usepackage{xcolor}

\def\Eq#1{Eq.~\eqref{#1}}

\def\Fig#1{Fig.~\ref{#1}}

\def\Sec#1{Sec.~\ref{#1}}
\def\App#1{App.~\ref{#1}}

\newcommand{\gettitle}{Solving Functional Renormalization Group Equations with Neural Networks}

\hypersetup{
	colorlinks,
	linkcolor={blue!75!black},
	citecolor={blue!75!black},
	urlcolor={blue!75!black},
	pdftitle={\gettitle},
	pdfauthor={},
	pdfkeywords={}
	{}
	bookmarksopen=true,
	bookmarksopenlevel=2,
	bookmarksnumbered=true
}

\begin{document}
\preprint{RIKEN-iTHEMS-Report-26}

\title{\gettitle}
\author{Yang-yang Tan}
\email{yangyang-tan@foxmail.com}
\affiliation{School of Physics, Dalian University of Technology, Dalian, 116024, China}
\affiliation{Department of Physics, Tsinghua University, Beijing 100084, China}
\affiliation{Institute for Physics of Intelligence, Graduate School of Science, The University of Tokyo, Bunkyo-ku, Tokyo 113-0033, Japan}

\author{Wei-jie Fu}
\email{wjfu@dlut.edu.cn}
\affiliation{School of Physics, Dalian University of Technology, Dalian, 116024, China}

\author{Lianyi He}
\email{lianyi@mail.tsinghua.edu.cn}
\affiliation{Department of Physics, Tsinghua University, Beijing 100084, China}

\author{Lingxiao Wang}
\email{lingxiao.wang@riken.jp}
\affiliation{RIKEN Center for Interdisciplinary Theoretical and Mathematical Sciences (iTHEMS), Wako, Saitama 351-0198, Japan}
\affiliation{Institute for Physics of Intelligence, Graduate School of Science, The University of Tokyo, Bunkyo-ku, Tokyo 113-0033, Japan}

\date{\today}

\begin{abstract}
We employ deep neural networks to represent the field derivative of the scale-dependent effective potential in the functional renormalization group (fRG) framework for nonperturbative quantum field theory. By embedding the fRG flow equations directly into the loss function, the network parameters are determined so as to provide a continuous and differentiable representation of the scale- and field-dependent effective potential without relying on precomputed training data. Focusing on the $O(N)$ scalar field theory within the local potential approximation at finite temperature, we demonstrate that this neural network representation accurately captures the renormalization group flow across symmetric, broken, and critical regimes. A key ingredient is a decomposition of the representation into an analytically known large-$N$ contribution and a learned finite-$N$ correction, which efficiently mitigates numerical stiffness associated with convexity restoration in the broken phase. The physics-driven solutions show excellent agreement with established finite-difference and discontinuous Galerkin methods. We further apply the same strategy to the Wilson--Fisher fixed-point equation in three dimensions, illustrating that neural network representations provide a unified framework for both scale-dependent flows and fixed-point problems. For fixed points, the large-field asymptotic form alone can replace the exact large-$N$ reference, while a composite small- and large-field ansatz further improves accuracy, extending the method to problems without an analytically solvable limit. Our results indicate that physics-driven deep learning offers a robust and flexible numerical tool for functional renormalization group studies.
\end{abstract}

\maketitle

\section{Introduction}

Understanding the non-perturbative dynamics of strongly correlated quantum systems remains one of the central challenges in modern theoretical physics. From the confinement of quarks and the phase diagram of quantum chromodynamics (QCD) at finite temperature and density, to quantum phase transitions and critical phenomena in condensed matter systems, many of the most fascinating physical phenomena occur in regimes where conventional perturbative expansions fail. The functional renormalization group (fRG)~\cite{Wetterich:1992yh} has emerged as a preeminent non-perturbative tool for addressing these challenges, implementing Kenneth Wilson's renormalization group philosophy~\cite{Wilson:1971bg, Wilson:1971dh, Wilson:1971dc, Wilson:1973jj} by systematically integrating out fluctuations scale by scale.

The fRG is built upon the Wetterich equation~\cite{Wetterich:1992yh} , an exact functional differential equation governing the scale dependence of the effective action $\Gamma_k[\phi]$. By introducing a momentum-dependent infrared regulator, the fRG smoothly interpolates between the classical action in the ultraviolet (UV) and the full quantum effective action in the infrared (IR). This approach has proven remarkably successful across diverse areas of physics, including the critical phenomena for strong correlated systems~\cite{Berges:2000ew,Canet:2006xu, Canet:2011wf, Mesterhazy:2013naa, Bluhm:2018qkf, Roth:2023wbp, Tan:2024fuq, Chen:2024lzz, Roth:2024rbi, Roth:2024hcu}, the QCD phase diagram~\cite{Braun:2014ata, Mitter:2014wpa, Rennecke:2015eba, Cyrol:2016tym, Cyrol:2017qkl, Cyrol:2017ewj, Fu:2019hdw, Braun:2020ada,Ihssen:2024miv,Pawlowski:2025jpg,Tan:2025bsv}, and the quantum gravity. One can refer to relevant reviews, e.g.,~\cite{Berges:2000ew, Pawlowski:2005xe, Braun:2011pp, Dupuis:2020fhh, Fu:2022gou} for more details.

Despite its conceptual elegance and broad applicability, practical calculations of fRG usually have to face significant computational challenges. The Wetterich equation involves infinite-dimensional functional spaces that necessitate systematic truncation schemes such as the derivative expansion or the vertex expansion. Even after truncation, the finite sets of flow equations are still challenging to be solved numerically in many cases.

In recent years, machine learning (ML) has emerged as a transformative tool across many branches of physics, offering new approaches to longstanding computational and theoretical challenges, especially in the QCD physics~\cite{Boehnlein:2021eym,Zhou:2023pti,Aarts:2025gyp}. Within the context of quantum field theory, ML techniques have found applications in lattice QCD simulations~\cite{Boyda:2022nmh,Cranmer:2023xbe,Aarts:2025gyp}, including generating gauge field configurations and accelerating Monte Carlo sampling. Diffusion models, connected to the stochastic quantization, have shown particular promise for lattice field theory simulations~\cite{Wang:2023exq,Wang:2023sry,Zhu:2024kiu,Zhu:2025pmw,Aarts:2026zzr}. A particularly powerful paradigm is physics-informed~\cite{RAISSI2019686,Karniadakis2021} or physics-driven learning~\cite{Aarts:2025gyp}, where physical laws and symmetries are directly embedded into the network architecture or loss function, enabling solutions that respect the underlying physics by construction.

In this work, we introduce a novel computational approach in which a neural network is used to represent the field derivative of the effective potential, with the fRG flow equations embedded directly into the loss function to determine the network parameters. This idea draws on the physics-informed neural network (PINN)~\cite{RAISSI2019686,Karniadakis2021}, which incorporates governing differential equations, boundary conditions, and initial conditions simultaneously into the loss function and has demonstrated remarkable success across scientific computing, including fluid dynamics, solid mechanics, etc. Recently, PINNs have been applied to solve the functional renormalization group equations on lattices~\cite{Yokota:2023czk}, or functional differential equations more generally~\cite{Miyagawa:2024yrw}, and the Dyson-Schwinger equations~\cite{Terin:2024iyy,CarmoTerin:2025pvs}. In contrast to the standard PINNs, our approach uses only the PDE residual of the fRG flow equation as the loss function, without imposing explicit boundary or initial conditions as additional constraints. In our case, we show that the neural network representations of the effective potential provide an effective approach for the fRG calculations.

We demonstrate our approach with the $O(N)$ scalar field theory within the local potential approximation (LPA), a paradigmatic model that captures the essential physics of spontaneous symmetry breaking and critical phenomena. The $O(N)$ model encompasses a wide range of universality classes, including the Ising model ($N=1$) for the liquid-gas and magnetic transitions, the XY model ($N=2$) for the superfluidity, the Heisenberg model ($N=3$) for the ferro(antiferro)-magnetism, and the linear sigma model ($N=4$) relevant for the chiral phase transition in two-flavor QCD. We solve the fRG flow equations at finite temperature across both the symmetric and broken phases, as well as the fixed point equation characterizing the Wilson-Fisher universality class. In all cases, the neural network results show excellent agreement with those obtained from well-established numerical methods, e.g., the Finite Difference Method and the Local Discontinuous Galerkin (LDG). More importantly, it offers a unified framework that naturally handles the stiffness and multi-scale structure of the flow.

This paper is organized as follows. In \Cref{sec:fRG} we provide a brief introduction about the fRG, including some theoretical basics, truncation schemes as well as some necessary notations. In \Cref{sec:ON} we present the fRG flow equations for the $O(N)$ model and the analytical large-$N$ solutions required for our methodology, with detailed derivations collected in \App{app:derivations}. In \Cref{sec:PINN} we discuss our neural network methodology, including the network architecture, loss function construction, and training strategies, followed by numerical results for the effective potential, physical observables, and the Wilson-Fisher fixed point. Finally, we summarize our findings and discuss future directions in \Cref{sec:conclusion}.

\section{The functional renormalization group method}
\label{sec:fRG}
The central insight underlying the fRG stems from Wilson's RG that the physics at different energy scales can be systematically separated and integrated out sequentially. In conventional quantum field theory, the path integral involves integration over all field configurations at all momentum scales simultaneously, making non-perturbative calculations intractable. The fRG resolves this by introducing a momentum-dependent infrared regulator that suppresses fluctuations below a running cutoff scale $k$, enabling the successive integration of quantum fluctuations from the UV to the IR scales.

The central object in fRG is the scale-dependent effective action $\Gamma_k[\phi]$, which interpolates between the classical action at the UV cutoff $\Lambda$ and the full quantum effective action in the IR limit:
\begin{align}
\Gamma_{k=\Lambda}[\phi] &= S[\phi] \quad \text{(classical action)}, \\
\Gamma_{k \to 0}[\phi] &= \Gamma[\phi] \quad \text{(full quantum effective action)}.
\end{align}
In this work we consider the case that $\phi$ represent the scalar field.

The effective action $\Gamma_k[\phi]$ is defined through a modified Legendre transform of the scale-dependent generating functional $W_k[J]$:
\begin{align}
\Gamma_k[\phi] + \Delta S_k[\phi] = \sup_J \left\{ \int J(x) \phi(x) - W_k[J] \right\},
\label{eq:Gamma_k_def}
\end{align}
where $\Delta S_k[\phi] = \frac{1}{2} \int \phi R_k \phi$ is the regulator which behaves like a mass term that suppresses IR fluctuations below the scale $k$.

The infrared regulator $R_k(p^2)$ plays a crucial role in the fRG formalism. It must satisfy several key properties:
\begin{itemize}
\item \textbf{IR suppression}: $R_k(p^2) > 0$ as $p^2 \to 0$ for fixed $k > 0$, suppressing low-momentum fluctuations.
\item \textbf{UV limit}: $R_k(p^2) \to \infty$ at the initial scale $k=\Lambda \to \infty$, which ensures the classical limit $\Gamma_{k=\Lambda}[\phi] = S[\phi]$.
\item \textbf{Physical limit}: $R_k(p^2) \to 0$ as $k \to 0$, where all quantum fluctuations are taken into account, yielding $\Gamma_{k \to 0}[\phi] = \Gamma[\phi]$ as the regulator is removed.
\end{itemize}

A particularly effective choice is the optimized Litim regulator~\cite{Litim:2000ci,Litim:2001up},
\begin{align}
R_k(p^2) = Z_{\phi,k} p^2 \left( \frac{k^2}{p^2} - 1 \right) \Theta(k^2 - p^2)\,, \label{eq:litim_regulator}
\end{align}
which provides optimal convergence properties~\cite{Litim:2001up} in the local potential approximation (LPA) and simplifies many calculations while maintaining the essential physics. The choice of regulator can be formulated as an optimization problem~\cite{Pawlowski:2005xe,Pawlowski:2015mlf}, which is also amenable to deep learning methods, and we hope to report our progress on this in the future.

The evolution of $\Gamma_k[\phi]$ with the RG scale is governed by the Wetterich equation:
\begin{align}
\partial_t \Gamma_k[\phi] = \frac{1}{2} \text{Tr} \left[ \partial_t R_k \left( \Gamma_k^{(2)}[\phi] + R_k \right)^{-1} \right]\,,
\label{eq:wetterich}
\end{align}
where $\partial_t \equiv k \partial_k$ denotes the RG time derivative with $t=\ln (k/\Lambda)$, $\Gamma_k^{(2)}[\phi]$ is the second functional derivative with respect to the field $\phi$, and $R_k$ is the infrared regulator function. The super-trace represents taking the trace over all internal indices as well as the momentum integration. The factor $\partial_t R_k$ ensures that only fluctuations around momentum scale $k$ contribute to the flow mostly, implementing Wilson's idea by integrating out fluctuations shell by shell in momentum space.

Although the Wetterich equation is formally one-loop exact, it remains impossible to obtain a full solution for the effective action except in some special cases, e.g., the zero-dimensional QFT. To render the equation tractable, systematic non-perturbative expansion methods have been developed to truncate the effective action, including the derivative expansion (DE) and the vertex expansion (VE). These approaches have been extensively tested across multiple physical systems, often capturing the essential physics without requiring high truncation orders. However, even after truncation, it remains challenging to solve the resulting flow equations, often necessitating additional approximation techniques for numerical feasibility.

In the DE, one expands the scale-dependent effective action $\Gamma_k[\phi]$ in powers of derivatives or momenta, which is valid for small momenta or long-wavelength physics~\cite{Balog:2019rrg,DePolsi:2020pjk}. This method has proved to be highly effective for studying low-energy QCD and critical phenomena in strongly correlated systems. The effective potential, comprised of the lowest-order terms in this expansion, is particularly significant as it encodes phase transition information and $n$-point vertex functions at vanishing external momenta to all orders. The flow equation for the effective potential manifests as a partial differential equation (PDE). Over the past two decades, various numerical methods have been developed to solve this equation, including expansions of the effective potential in Taylor series~\cite{Litim:2002cf} or Chebyshev polynomials~\cite{Borchardt:2015rxa,Borchardt:2016pif,Chen:2021iuo}. More recent developments involve recasting the equation in conservative form and applying modern PDE solvers such as the finite element method~\cite{Grossi:2019urj,Sattler:2024ozv} and the finite volume method~\cite{Zorbach:2024rre}. However, these approaches become increasingly complex and computationally intensive when extended to higher-dimensional field spaces.

The VE offers an alternative to DE by systematically expanding the effective action in powers of the field, thereby capturing the full momentum dependence. Although we do not discuss the VE in this work, we briefly mention it here for completeness. This approach is widely employed in studies of Yang-Mills theory~\cite{Cyrol:2016tym}, QCD~\cite{Fu:2022uow,Fu:2025hcm}, and quantum gravity, where propagators and higher-order vertices exhibit non-trivial momentum structures. Additionally, the vertex expansion is crucial for investigating spectral functions in real-time fRG~\cite{Tan:2021zid}. The corresponding flow equations often take the form of functional differential-integral equations, for which there is no well-established numerical method yet. The conventional approach involves discretizing $n$-point functions on a momentum grid; however, even a simple four-point vertex function may have six degrees of freedom, and resolving its full momentum dependence would demand over a billion grid points ($32^6$), making traditional discretization methods computationally prohibitive. Further momentum-space approximations are typically required to render the problem numerically tractable. Neural network representations are well suited for such high-dimensional problems, and we plan to explore their application to the VE in future work.

\section{The fRG flows in the $O(N)$ model}
\label{sec:ON}
The $O(N)$ model describes $N$ real scalar fields $\phi_a$ ($a = 1, \ldots, N$) with global $O(N)$ symmetry, serving as a paradigmatic system for understanding critical phenomena and phase transitions. In this work, we focus on the finite-temperature dynamics with purely dissipative relaxation (Model A in the Hohenberg-Halperin classification~\cite{Hohenberg:1977ym}), which provides a minimal description of thermal phase transitions and critical dynamics in fRG. In this section, we present the flow equations and their analytical large-$N$ solutions essential for our neural network methodology. Detailed derivations are provided in \App{app:derivations}.

\subsection{Finite-temperature flow equation}
\label{sec:flow_eq}
In this work, we adopt the local potential approximation (LPA), for which the effective action takes the form
\begin{align}
\Gamma_k[\phi]=\int \mathrm{d}^d x\left[\frac{1}{2} Z_{\phi, k}\left(\partial_\mu \phi_a\right)^2+V_k(\rho)+\mathcal{O}(\partial^2)\right]\,,
\end{align}
where $\rho = \frac{1}{2}\phi_a \phi_a$ is the $O(N)$-invariant field combination, $Z_{\phi,k}$ is the wave-function renormalization with $Z_{\phi,k}=1$ for LPA, and $V_k(\rho)$ is the scale-dependent effective potential. We use the standard notation: $t\equiv\ln(k/\Lambda)$ for the RG time, $\eta\equiv-\partial_t\ln Z_{\phi,k}$ for the anomalous dimension, and primes for derivatives with respect to field.

For finite-temperature systems with purely dissipative dynamics (Model A~\cite{Hohenberg:1977ym}), the flow equation of the effective potential reads (see \App{app:finite_T} for derivation):
\begin{align}
	\partial_t V_k(\rho)=\mathscr{C}T k^d\left[\frac{1}{1+\bar{m}_{\sigma, k}^2}+\frac{N-1}{1+\bar{m}_{\pi, k}^2}\right]\,,
	\label{eq:flow_potential_modelA}
\end{align}
where the dimensionless curvature masses of the longitudinal ($\sigma$) and transverse ($\pi$) modes are defined as
\begin{align}
	\bar{m}_{\sigma, k}^2 &= \frac{V_k^{\prime}+2 \rho V_k^{(2)}}{Z_{\phi, k} k^2}\,,\nonumber\\[2ex]
	\bar{m}_{\pi, k}^2 &= \frac{V_k^{\prime}}{Z_{\phi, k} k^2}\,.
\end{align}
The constant, which reads
\begin{align}
	\mathscr{C} \equiv \frac{1}{(4 \pi)^{d / 2} \Gamma(d / 2)}\big[(2-\eta)/d+\eta/(d+2)\big]\,,
\end{align}
is a geometric prefactor originating from the internal momentum integration and carries the explicit dependence on the spatial dimension $d$ in the flow equation. The factor of $T$ reflects the dominance of thermal fluctuations, which controls the overall magnitude of the fRG flow.

For numerical implementation, we rescale the field and potential using the critical temperature $T_c$ as $\hat{\rho}=\rho/(\Lambda^{d-2} T_c)$ and $\hat{V}_k=V_k/(\Lambda^d T_c)$. Substituting $k=\Lambda e^t$ with $t\leq 0$ and dividing numerator and denominator of each propagator by $k^2$ yields the rescaled flow equation:
\begin{align}
	\partial_t \hat{V}_k(\hat{\rho})= &\mathscr{C}\frac{T}{T_c}\mathrm{e}^{(d+2)t}\Bigg[\frac{1}{\mathrm{e}^{2t}+\hat{m}_{\sigma, k}^2}+\frac{N-1}{\mathrm{e}^{2t}+\hat{m}_{\pi, k}^2}\Bigg]\,,
	\label{eq:flow_potential_rescale}
\end{align}
where $\hat{m}_{\sigma, k}^2 \equiv \hat{V}'_k+2 \hat{\rho} \hat{V}''_k$ and $\hat{m}_{\pi, k}^2 \equiv \hat{V}'_k$. This is the primary equation whose solution is represented by our neural network for the finite-temperature results in \Sec{sec:PINN}.

\subsection{Large-N analytic solution}
\label{sec:largeN}
In the large-$N$ limit, the sigma mode becomes suppressed relative to the $(N-1)$ pion modes, and the flow equation \eqref{eq:flow_potential_modelA} reduces to:
\begin{align}
	\partial_t V_k(\rho)=\mathscr{C}T k^d\frac{N-1}{1+\bar{m}_{\pi, k}^2}\,.
	\label{eq:flow_potential_modelA_largeN}
\end{align}

This first-order PDE can be solved exactly using the method of characteristics (see \App{app:characteristics}). The result is an implicit solution for $U_k(\rho) \equiv V'_k(\rho)$:
\begin{align}
	\rho_k &= \rho_\Lambda+\frac{1}{2U}\left(\frac{k^{d+2}}{k^2+U}-\frac{\Lambda^{d+2}}{\Lambda^2+U}\right)\nonumber\\[2ex]
	&+\frac{d}{2(d+2)U^2}\left[\Lambda^{d+2}\, _2F_1\left(1,\frac{d+2}{2};\frac{d+4}{2};-\frac{\Lambda^2}{U}\right)\right.\nonumber\\[2ex]
	&\left. -k^{d+2} \, _2F_1\left(1,\frac{d+2}{2};\frac{d+4}{2};-\frac{k^2}{U}\right) \right]\,,
	\label{eq:largeN_solution}
\end{align}
where $_2F_1$ denotes the hypergeometric function. Along each characteristic, $U$ stays constant, being equal to its UV value, while the field coordinate $\rho_k$ evolves with the RG scale.

Although the large-$N$ limit does not capture all quantitative aspects of the finite-$N$ theory, it preserves the essential qualitative structure of the renormalization group flow. In particular, it correctly describes symmetry restoration, the emergence of a flat effective potential in the broken phase, and the approach to criticality. In the context of numerical RG calculations, the analytic large-$N$ solution therefore serves as a natural baseline. This observation motivates the strategy adopted in \Sec{sec:PINN}, where we decompose the full finite-$N$ solution into an analytically known large-$N$ contribution and a numerically learned correction. By isolating the dominant large-$N$ behavior, the remnant correction will reduce the stiffness induced by convexity restoration. As we demonstrate below, this decomposition substantially stabilizes the neural network training and enhances accuracy across all temperature regimes.
\begin{figure*}[htbp!]
	\centering
	\includegraphics[width=0.5\textwidth]{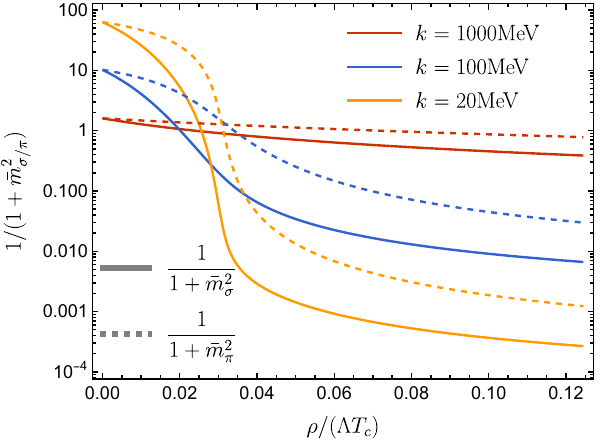}\includegraphics[width=0.48\textwidth]{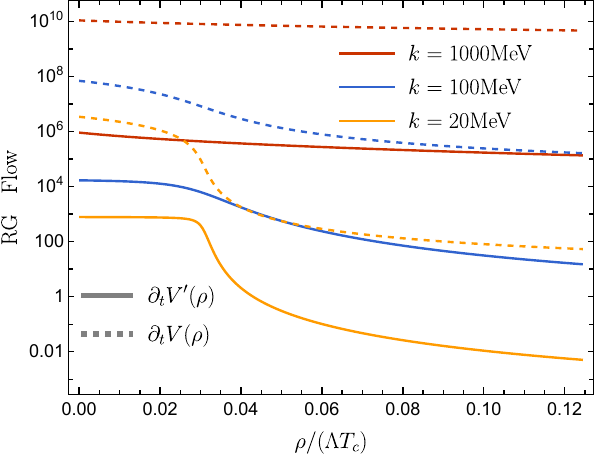}
	\caption{Left panel: The dimensionless sigma and pion propagators $1/(1+\bar{m}_{\sigma}^2)$ (solid lines) and $1/(1+\bar{m}_{\pi}^2)$ (dashed lines) as functions of the rescaled field $\hat{\rho}=\rho/(\Lambda T_c)$ at different RG scales $k=1000, 100, 20\,\mathrm{MeV}$. Right panel: The RG flow of the effective potential $\partial_t V(\rho)$ (dashed lines) and its field derivative $\partial_t V'(\rho)$ (solid lines) at the same RG scales. Both panels correspond to the $O(4)$ model at $T=100\,\mathrm{MeV}$ in the broken phase.}
    \label{fig:sigma-pi-propagator-V-Vp-flow}
\end{figure*}
\subsection{Fixed point equation}
\label{sec:fixedpoint_eq}
For studying universal critical behavior, we also solve the Wilson-Fisher fixed point equation. This is derived from the zero-temperature Euclidean flow equation (see \App{app:derivations}), which can also be derived from \Eq{eq:flow_potential_modelA} by proper rescaling. At the fixed point, the dimensionless potential $u(\bar{\rho})=k^{-d} V_k(\rho)$ becomes scale-invariant, satisfying $\partial_t u = 0$. With $\bar{\rho}=k^{-(d-2)} Z_{\phi, k} \rho$, the fixed point equation reads:
\begin{align}
	0 = & -d u+(d-2+\eta) \bar{\rho} u^{\prime} +\mathscr{C}\left[\frac{1}{1+u^{\prime}+2 \bar{\rho} u^{(2)}}+\frac{N-1}{1+u^{\prime}}\right]\,.
	\label{eq:fixed_point}
\end{align}

In the large-$N$ limit, this equation can be solved exactly in an implicit form:
\begin{align}
	\bar{\rho} = \frac{\mathscr{C}(N-1)}{d-2+\eta}{}_2F_1\left(2,\frac{d-2+\eta}{\eta-2};\frac{d-4+2\eta}{\eta-2};-u_*'(\bar{\rho})\right)\,.
	\label{eq:fixed_point_largeN}
\end{align}
Similar with the finite-temperature flow, our neural network learns the finite-$N$ correction $\Delta u' = u' - u'_{\mathrm{LN}}$ with respect to this large-$N$ baseline, such that one can efficiently avoid the convergence of solutions towards the trivial Gaussian fixed point.

\section{Neural network representation of the effective potential}
\label{sec:PINN}
Before presenting our neural network approach, we first illustrate the numerical challenges inherent in solving the fRG flow equations. \Fig{fig:sigma-pi-propagator-V-Vp-flow} displays field-dependent fRG flow and its components for the $O(4)$ model at $T=100\,\mathrm{MeV}$ in the broken phase. The left panel shows the dimensionless sigma and pion propagators, $1/(1+\bar{m}_{\sigma}^2)$ (solid lines) and $1/(1+\bar{m}_{\pi}^2)$ (dashed lines), as functions of the rescaled field $\hat{\rho}=\rho/(\Lambda T_c)$ at three representative RG scales. At the UV scale ($k=1000\,\mathrm{MeV}$), both propagators are of order unity and vary smoothly with the field. As the RG scale decreases, both propagators develop rapid variations spanning several orders of magnitude within a narrow field range around the potential minimum $\rho_0$. This dramatic field dependence in the small-field region is the primary source of numerical stiffness in the fRG flow equations.

The right panel of \Fig{fig:sigma-pi-propagator-V-Vp-flow} shows the corresponding RG flows $\partial_t V(\rho)$ (dashed lines) and $\partial_t V'(\rho)$ (solid lines). At the UV scale ($k=1000\,\mathrm{MeV}$), both flows are large but exhibit relatively uniform magnitude across the field range. As $k$ decreases, although the overall magnitude of the flows decreases due to the $k^d$ prefactor in the flow equation, the ratio between the maximum and minimum magnitudes within each RG scale increases dramatically, spanning from $\mathcal{O}(10^{-2})$ to $\mathcal{O}(10^{3})$ and from $\mathcal{O}(10^2)$ to $\mathcal{O}(10^{7})$ at $k=20\,\mathrm{MeV}$ for the effective potential and its field derivative, respectively. This enormous dynamic range, concentrated around the potential minimum $\rho_0$ where the convexity restoration occurs, constitutes the primary numerical challenge. Standard ODE/PDE solvers require extremely small step sizes to resolve these rapid variations, leading to prohibitive computational costs.

In the broken phase, the effective potential $V_k(\rho)$ at finite $k$ is non-convex in the region $\rho < \rho_0$, reflecting the underlying spontaneous symmetry breaking. However, the full quantum effective action, obtained as the Legendre transform of the Schwinger functional, must be a convex function of the field. As shown in \Eq{eq:Gamma_k_def}, while $\Gamma_k[\phi]$ itself need not be convex at finite $k$, the combination $\Gamma_k[\phi]+\Delta S_k[\phi]$ remains convex throughout the RG flow. This implies $\Gamma^{(2)}_k[\phi]+R_k\geq 0$, which for the $O(N)$ model with the Litim regulator guarantees positivity of the denominators in the flow equation, i.e., $1+\bar{m}_\sigma^2\geq 0$ and $1+\bar{m}_\pi^2\geq 0$. As $k\to 0$ and $\Delta S_k[\phi] \to 0$, quantum fluctuations flatten the potential in the non-convex region gradually, ultimately restoring the full convexity.

\subsection{Neural network architecture}

\begin{figure}[htbp!]
	\centering
	\includegraphics[width=0.48\textwidth]{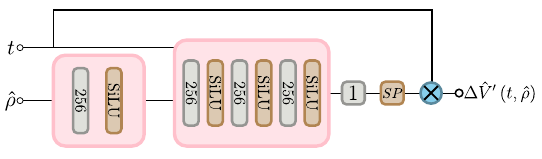}
	\caption{Neural network architecture for learning the fRG flow. The network takes as input the RG scale parameter $t$ and the rescaled field variable $\hat{\rho}$, and outputs the finite-$N$ correction to the field derivative of the effective potential, $\Delta\hat{V}'(t,\hat{\rho}) = \hat{V}'(t,\hat{\rho}) - \hat{V}'_{\mathrm{LN}}(t,\hat{\rho})$. The architecture consists of two sub-networks, a field encoding network and an output network with concatenated inputs.}
	\label{fig:NN_architecture}
\end{figure}

The neural network architecture employed in this work, illustrated in \Fig{fig:NN_architecture}, is specifically designed to handle the RG evolution of the field-dependent effective potential and is easy to extend to multi-field dependence in future research. The network predicts the finite-$N$ correction to the field derivative of the effective potential:
\begin{align}
\Delta\hat{V}'(t,\hat{\rho}) = \hat{V}'(t,\hat{\rho}) - \hat{V}'_{\mathrm{LN}}(t,\hat{\rho})\,,
\end{align}
where $\hat{V}'_{\mathrm{LN}}(t,\hat{\rho})$ denotes the analytically known large-$N$ solution obtained via the method of characteristics described in \Sec{sec:largeN}. The architecture consists of two interconnected sub-networks that process the field and RG scale information separately.

The network takes two inputs, the RG scale parameter $t = \ln(k/\Lambda)$ and the rescaled field variable $\hat{\rho} = \rho/(\Lambda^{d-2} T_c)$. The rescaled field $\hat{\rho}$ is firstly processed through a dedicated encoding network consisting of a single fully connected layer with 256 hidden units. This layer employs the Sigmoid Linear Unit (SiLU) activation function, defined as $\text{SiLU}(x) = x \cdot \sigma(x)$ where $\sigma$ is the sigmoid function. The SiLU activation was chosen for its non-monotonic properties that have proven effective in capturing the complex field dependence of the effective potential. For multi-field dependence in future studies, one can use a more sophisticated field encoding network. The 256-dimensional field representation is concatenated with the RG scale parameter $t$, yielding a 257-dimensional feature vector. This combined representation is processed through a deep network consisting of three fully connected layers, each with 256 hidden units and SiLU activations, followed by a final output layer. The final layer employs a Softplus activation function, $\text{Softplus}(x) = \ln(1 + e^x)$, ensuring strictly positive outputs. Crucially, this output is multiplied by the RG scale parameter $t$, implementing the \textit{boundary condition} $\Delta\hat{V}'(t=0,\hat{\rho}) = 0$ exactly. This multiplicative structure ensures that the network output vanishes at the UV scale, since we set the large-$N$ solution to have the same \textit{initial condition} as the finite-$N$ one. The network contains approximately 198,000 trainable parameters, distributed primarily across the hidden layers.

\begin{figure*}[t]
	\centering
	\hspace{-1em}\includegraphics[width=0.48\textwidth]{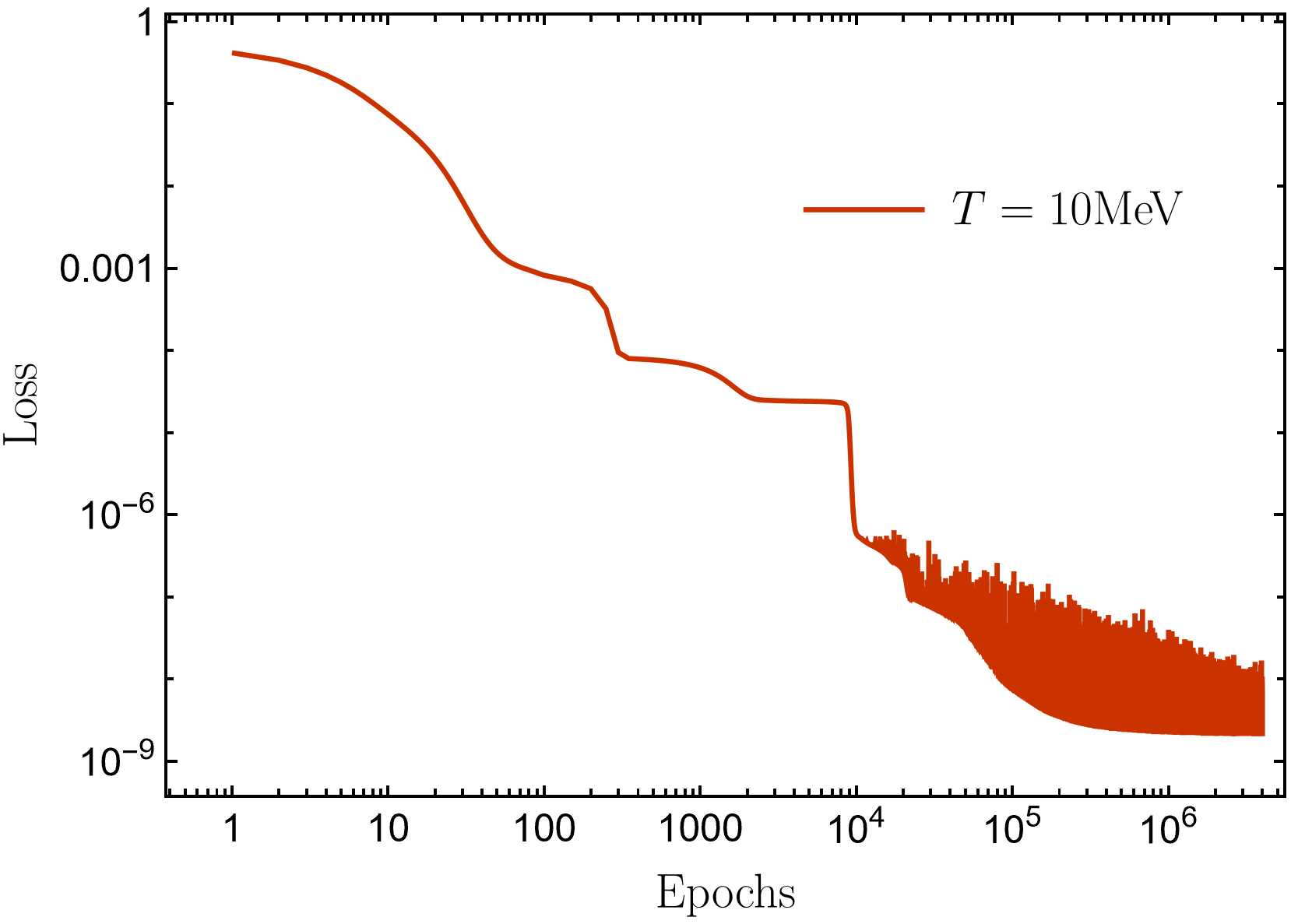}\includegraphics[width=0.48\textwidth]{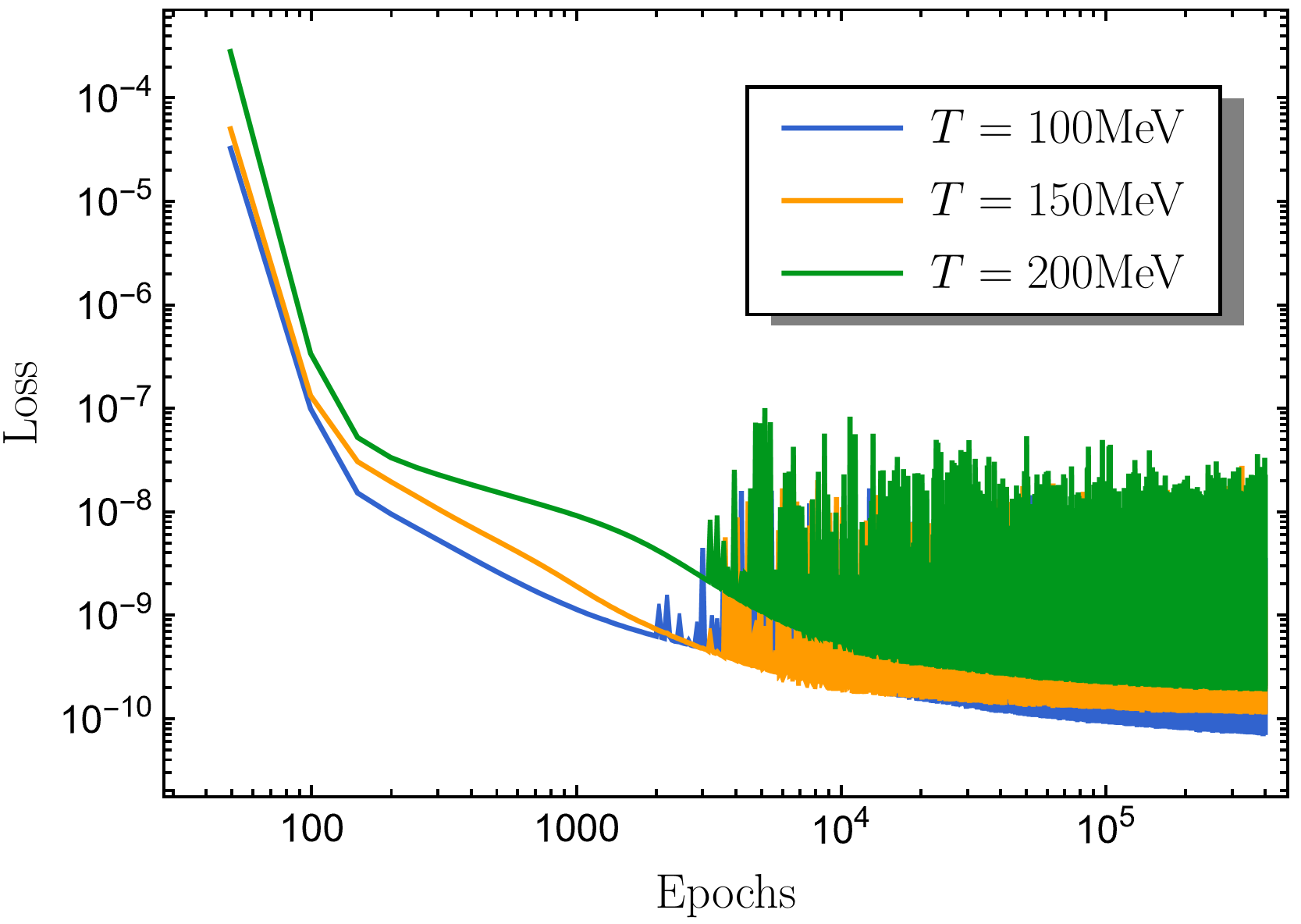}
	\caption{Training loss convergence for the $O(4)$ model. Left panel: Loss evolution at $T=10\,\mathrm{MeV}$ in the deeply broken phase over $4 \times 10^6$ epochs, displaying characteristic staircase-like multi-stage descent followed by a fluctuating refinement phase with loss around $10^{-9}$. Right panel: Loss evolution for transfer learning at $T=100, 150, 200\,\mathrm{MeV}$ using pre-trained weights from a model at $T=120\,\mathrm{MeV}$ as initialization. The transfer learning dramatically accelerates convergence, with all temperatures reaching loss values around $10^{-9}$ to $10^{-10}$ within $\sim 5\times 10^5$ epochs.}
	\label{fig:loss_O4}
\end{figure*}

The training of the neural network is guided by a loss function that enforces the fRG flow equation as a physics constraint. In contrast to traditional data-driven approaches, our method does not require pre-computed solutions but instead learns directly from the differential equation structure.

The loss function consists solely of the physics residual:
\begin{align}
	L(\theta) = \frac{1}{N_{\text{grid}}} \sum_{i,j} \left| \mathcal{R}_{ij}(\theta) \right|^2\,,
\end{align}
where $\mathcal{R}_{ij}(\theta)$ is the residual of the discretized flow equation evaluated at grid point $(i,j)$ corresponding to $(\hat{\rho}_i, t_j)$, and $N_{\text{grid}}$ is the total number of interior grid points.

The residual is computed using a finite difference discretization:
\begin{align}
	\mathcal{R}_{ij}(\theta) = \partial_t \hat{V}'_{\theta}(\hat{\rho}_i, t_j) - \mathcal{F}\left[\hat{V}'_{\theta}, \hat{V}''_{\theta},\hat{V}'''_{\theta}, \hat{\rho}_i, t_j\right]\,,
\end{align}
where $\mathcal{F}$ represents the right-hand side of the flow equation \eqref{eq:flow_potential_rescale}, and the derivatives are computed using three-point finite difference stencils. Spatial derivatives with respect to $\hat{\rho}$ are computed using central differences with second-order accuracy, while temporal derivatives can employ either central differences or backward Euler discretization for enhanced stability. This approach reduces memory consumption by approximately a factor of 3--4 and is significantly faster compared to automatic differentiation. The finite difference scheme may even lead to better numerical stability than automatic differentiation, as discussed in Ref.~\cite{Chiu_2022}. The training data consists of a structured grid spanning $\hat{\rho} \in [0, \hat{\rho}_{\max}]$ and $t \in [t_{\mathrm{IR}}, 0]$. For the results presented here, we employ grids with typical resolutions of $N_{\rho} \times N_t = 201 \times 501$, chosen to adequately resolve the sharp features in the broken phase while maintaining computational tractability. For the deeply broken phase, i.e., $T=10\,\mathrm{MeV}$, we employ grids with resolutions of $N_{\rho} \times N_t = 2001 \times 501$. This grid-based approach is suitable for our two-dimensional input space and ensures uniform coverage of the domain. However, when dealing with high-dimensional input spaces, quasi-random sampling methods may be more suitable. No explicit boundary conditions are imposed in the field direction, allowing the network to naturally determine the asymptotic behavior based on the physics encoded in the flow equation. This approach avoids potential artifacts from artificial boundary constraints that traditional numerical methods impose.

To ensure stable training in regions where the flow equation becomes nearly singular, we introduce a small regularization parameter $\epsilon = 10^{-13}$ in the denominators:
\begin{align}
	\frac{1}{1 + \bar{m}^2} \to \frac{1}{1 + \bar{m}^2 + \epsilon}\,.
\end{align}
This regularization is orders of magnitude smaller than the physical scales and does not affect the accuracy of the solution while preventing numerical instabilities during training.

The implementation leverages modern deep learning frameworks and optimization techniques tailored for scientific computing applications. The network is implemented using \textit{PyTorch} and the \textit{PhysicsNEMO} library~\cite{PhysicsNeMo:2023}, which provides specialized layers and utilities for scientific computing with neural networks. We use \textit{PyTorch Lightning}~\cite{PyTorchLightning:2019} to manage efficient GPU utilization and automatic checkpointing. All computations are performed in single precision (float32) by default. For the deeply broken case at $T=10\,\mathrm{MeV}$, the late stage of training is further refined in double precision (float64) to improve numerical stability and better resolve the sharp structure near the potential minimum.

\subsection{Numerical results}
We present numerical results for the $O(4)$ model at various temperatures to demonstrate the effectiveness of our neural network approach for learning fRG flows in the broken phase. The training is performed on a domain with RG time $t \in [-4, 0]$ (with $t=\ln(k/\Lambda)$), corresponding to the RG scale range $k \in [k_{\mathrm{IR}}, k_{\mathrm{UV}}]$ with $k_{\mathrm{UV}} = 1\,\mathrm{GeV}$ and $k_{\mathrm{IR}}\approx 20\mathrm{MeV}$. The rescaled field domain is chosen as $\hat{\rho} \in [0, 0.15]$.

\subsubsection{Training convergence and optimization strategy}

The optimization of the neural network parameters employs the Adam optimizer with a fixed learning rate of $5 \times 10^{-4}$. We find that the training results are largely insensitive to the choice of learning rate within a reasonable range.

\Fig{fig:loss_O4} illustrates the convergence characteristics of our training procedure. The left panel shows the training loss at $T=10\,\mathrm{MeV}$ in the deeply broken phase, where the network is trained for $4 \times 10^6$ epochs. The loss exhibits a characteristic staircase-like multi-stage descent, with rapid drops occurring around epochs $10^1$, $10^2$, and $10^4$ as the network progressively learns different scales of the flow structure. After approximately $10^5$ epochs, the training enters a fluctuating refinement phase where the loss oscillates around $10^{-9}$. This extended training is crucial for accurately capturing the subtle features around the effective potential minimum, which is important for the extraction of physical observables.

To improve the training efficiency across different temperature regimes, we employ a transfer learning strategy. We use the network parameters from a trained model at $T=120\,\mathrm{MeV}$ as initialization for training at other temperatures. As shown in the right panel of \Fig{fig:loss_O4}, this strategy dramatically accelerates convergence compared to training from scratch. All three temperatures ($T=100, 150, 200\,\mathrm{MeV}$) achieve loss values around $10^{-9}$ to $10^{-10}$ within $5\times 10^5$ epochs. This demonstrates that the learned representation generalizes well across different temperature regimes, significantly reducing the computational cost for systematic temperature scans.

\subsubsection{Effective potential in the broken phase}
\begin{figure*}[t]
	\centering
	\hspace{-1em}\includegraphics[width=0.48\textwidth]{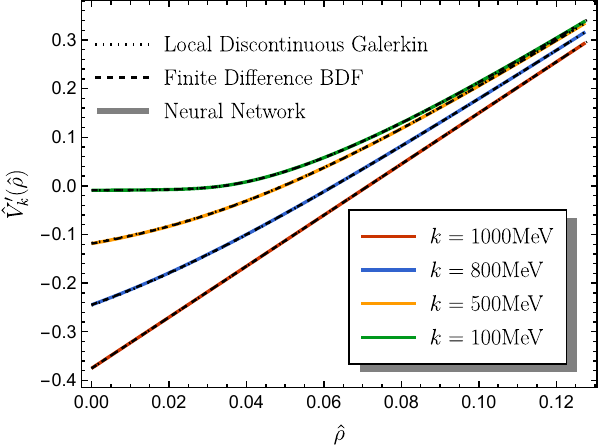}\includegraphics[width=0.48\textwidth]{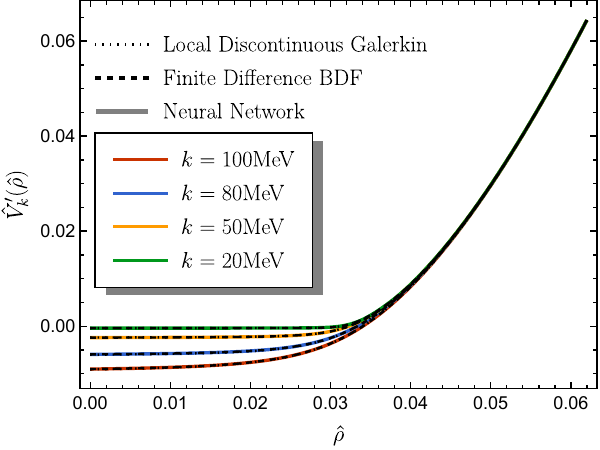}
	\caption{Field derivative of the rescaled effective potential $\hat{V}'_k(\hat{\rho})$ as a function of the rescaled field $\hat{\rho}$ for the $O(4)$ model at $T=100\,\mathrm{MeV}$. Left panel: Evolution at high RG scales ($k = 1000, 800, 500, 100\,\mathrm{MeV}$). Right panel: Evolution at low RG scales ($k = 100, 80, 50, 20\,\mathrm{MeV}$), where convexity restoration flattens $\hat{V}'_k$ in the small-field region. The neural network results (solid gray lines) are compared with the Local Discontinuous Galerkin (LDG) method (dotted lines) and Finite Difference BDF solver (dashed lines), showing excellent agreement across all RG scales.}
	\label{fig:T100_solutions}
\end{figure*}
\begin{figure*}[t]
	\centering
	\hspace{-1em}\includegraphics[width=0.48\textwidth]{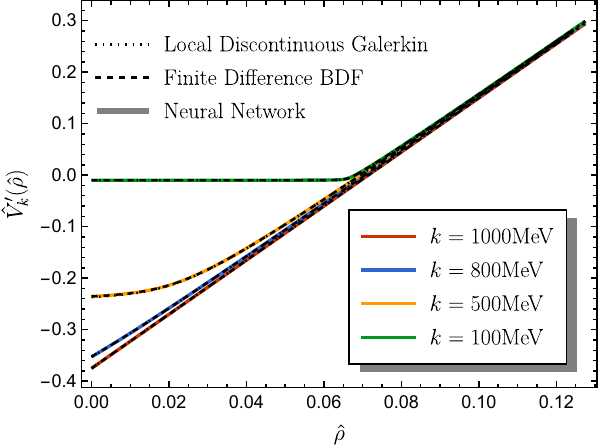}\includegraphics[width=0.48\textwidth]{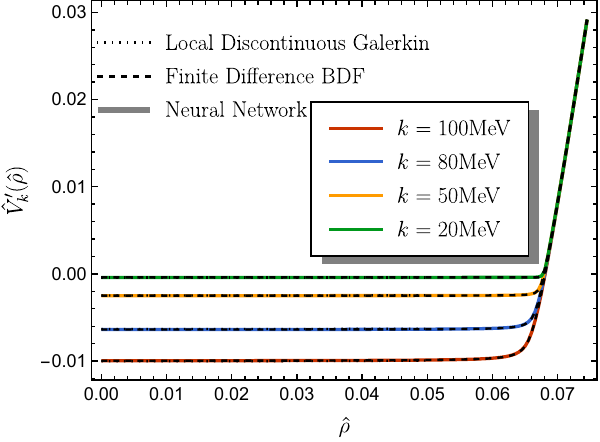}
	\caption{Field derivative of the rescaled effective potential $\hat{V}'_k(\hat{\rho})$ as a function of the rescaled field $\hat{\rho}$ for the $O(4)$ model in the deeply broken phase at $T=10\,\mathrm{MeV}$. Left panel: Evolution at high RG scales ($k = 1000, 800, 500, 100\,\mathrm{MeV}$). Right panel: Evolution at low RG scales ($k = 100, 80, 50, 20\,\mathrm{MeV}$), showing the convexity restoration in the small-field region. The neural network results (solid gray lines) are compared with the Local Discontinuous Galerkin method (dotted lines) and Finite Difference BDF solver (dashed lines), demonstrating excellent agreement across all RG scales.}
	\label{fig:T10_solutions}
\end{figure*}
\begin{figure*}[t]
	\centering
	\hspace{-1em}\includegraphics[width=0.48\textwidth]{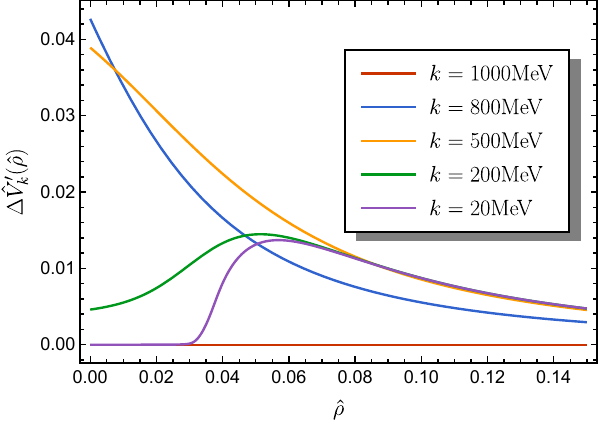}\includegraphics[width=0.48\textwidth]{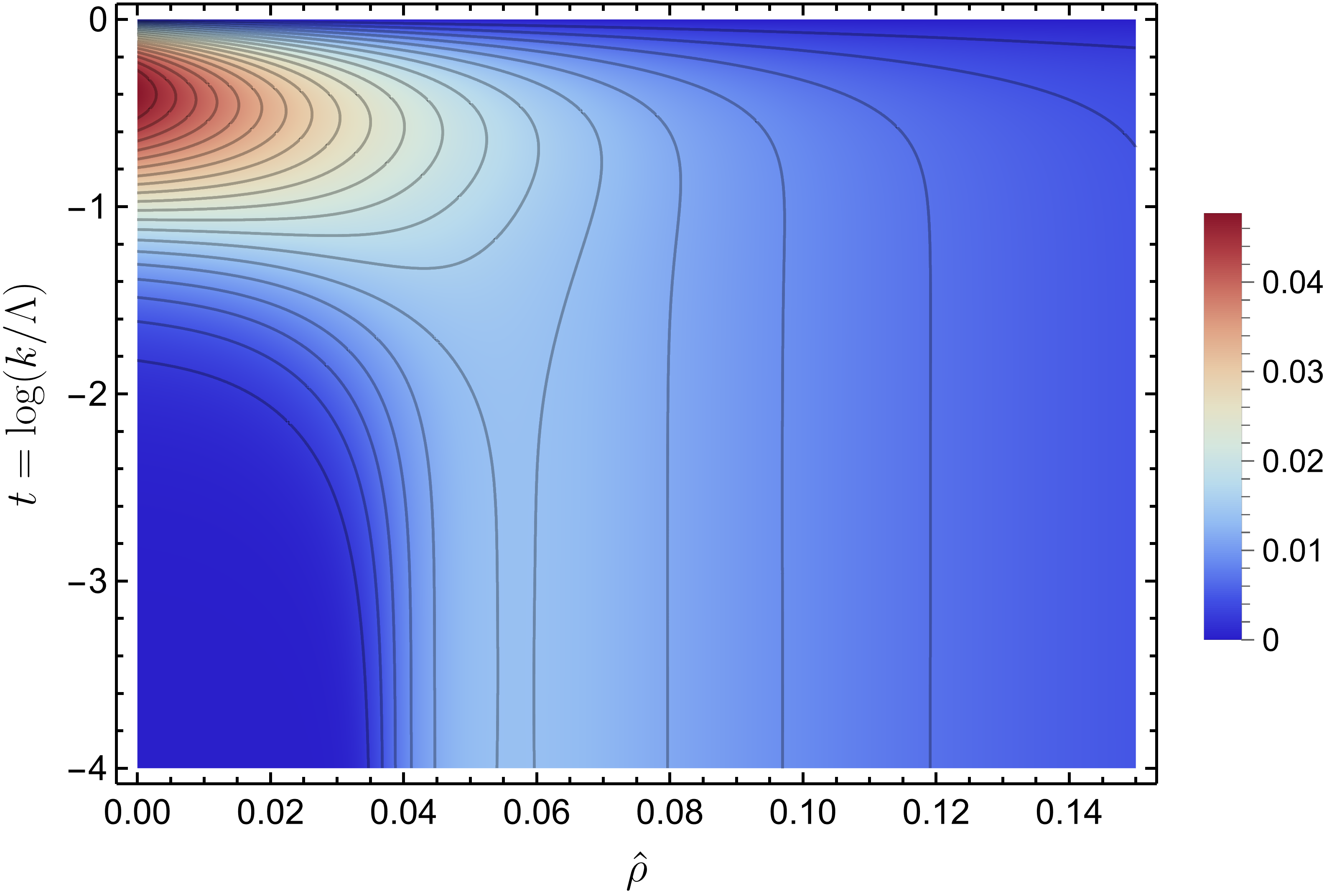}
	\caption{Neural network output $\Delta\hat{V}'_k(\hat{\rho}) = \hat{V}'_k(\hat{\rho}) - \hat{V}'_{\mathrm{LN},k}(\hat{\rho})$ representing the finite-$N$ correction learned by the neural network for the $O(4)$ model at $T=100\,\mathrm{MeV}$. Left panel: Field dependence at different RG scales $k = 1000, 800, 500, 200, 20\,\mathrm{MeV}$. At the UV scale ($k = 1000\,\mathrm{MeV}$), the correction vanishes by construction. Right panel: Heat map of $\Delta\hat{V}'(t,\hat{\rho})$ over the full $(t, \hat{\rho})$ domain, where $t = \ln(k/\Lambda)$. The smooth color gradients demonstrate the network's continuous interpolation across the training domain, with the largest finite-$N$ corrections (red regions) concentrated at small $\hat{\rho}$ and intermediate RG scales.}
	\label{fig:deltaV}
\end{figure*}
\Fig{fig:T100_solutions} presents the field derivative of the rescaled effective potential $\hat{V}'_k(\hat{\rho})$ for the $O(4)$ model at $T=100\,\mathrm{MeV}$, which lies below the critical temperature $T_c \approx 150\,\mathrm{MeV}$. The left panel displays the evolution at high RG scales ($k = 1000, 800, 500, 100\,\mathrm{MeV}$), where the potential derivative varies smoothly from negative values at small $\hat{\rho}$ to positive values at large $\hat{\rho}$, crossing zero at the potential minimum. At these scales, the system retains memory of the UV initial conditions, and the flow has not yet developed the strong nonlinearities characteristic of the broken phase.

As the RG scale decreases toward the infrared (right panel, $k = 100, 80, 50, 20\,\mathrm{MeV}$), the potential derivative becomes increasingly flat in the small-field region ($\hat{\rho} \lesssim 0.03$), approaching zero. This flattening is a direct manifestation of the convexity restoration mechanism, where quantum fluctuations progressively smooth out the non-convex structure of the effective potential.

We validate our neural network results (solid gray lines) against two established numerical methods, the Local Discontinuous Galerkin (LDG) method~\cite{Sattler:2024ozv} (dotted lines) and the Finite Difference solver with backward differentiation formula (BDF) for stiff equations (dashed lines). The agreement between all three methods is excellent across the entire field range and all RG scales, with absolute errors remaining below $4\times 10^{-5}$ even in the challenging small-field region where convexity restoration occurs or near the potential minimum.

The small-field region poses the greatest challenge for the neural network due to the rapid variations and near-singular behavior of the flow equation during convexity restoration. Our network architecture, which separates the analytically known large-$N$ contribution from the learned finite-$N$ correction, proves essential for achieving this accuracy. The large-$N$ solution captures the dominant physics of convexity restoration, allowing the neural network to focus on learning the comparatively smoother finite-$N$ corrections.

Beyond serving as a validation benchmark, \Fig{fig:T100_solutions} also provides an intuitive geometric picture of the RG evolution. In the broken phase, the running minimum $\hat{\rho}_0(k)$ corresponds to the location where $\hat{V}'_k(\hat{\rho})=0$, i.e., where the curves cross zero. As $k$ decreases, the flat region at small $\hat{\rho}$ expands and the zero-crossing point shifts toward smaller fields, reflecting the flow of the order parameter toward its infrared value. This geometric feature underlies the observable flows shown below. \Fig{fig:rho0_flow} tracks the motion of $\hat{\rho}_0(k)$, while \Fig{fig:msig2_flow} probes the local curvature at the running minimum through $\hat{V}''(\hat{\rho}_0)$.

\Fig{fig:T10_solutions} shows the same field derivative $\hat{V}'_k(\hat{\rho})$ in the deeply broken phase at $T=10\,\mathrm{MeV}$, where convexity restoration becomes more pronounced. The UV initial condition is identical to the $T=100\,\mathrm{MeV}$ case. Due to the overall $T$ prefactor in the flow equation \eqref{eq:flow_potential_modelA}, the flow magnitude is reduced at lower temperature, causing the potential minimum to shift less toward smaller fields. However, the convexity restoration mechanism remains active, driving the propagator denominators toward zero in the small-field region and generating large flows there. Consequently, while the RG evolution is nearly frozen for $\hat{\rho}\gtrsim \hat{\rho}_0$, it remains substantial for $\hat{\rho}\lesssim \hat{\rho}_0$.

As the scale is lowered (right panel), the solution develops a small-field plateau where $\hat{V}'_k$ remains nearly constant for $0\le \hat{\rho}\lesssim \hat{\rho}_0(k)$, with a sharp crossover around $\hat{\rho}_0(k)$ connecting to the rising branch at larger fields. As $k\to 0$, this plateau region expands and $\hat{V}'_k$ approaches zero, ultimately restoring full convexity. Compared with \Fig{fig:T100_solutions}, the crossover occurs at larger $\hat{\rho}$ and becomes significantly sharper, illustrating the increased stiffness of the flow deep in the broken phase. The excellent agreement between the neural network (solid gray) and the LDG/BDF benchmarks demonstrates that the network accurately resolves both the near-flat region and the steep gradients around $\hat{\rho}_0(k)$ without introducing spurious oscillations.

To gain deeper insight into the network's learned representation, we examine the neural network output $\Delta\hat{V}'_k(\hat{\rho})$ directly in \Fig{fig:deltaV}. This quantity represents the finite-$N$ correction to the analytically known large-$N$ solution.

The left panel of \Fig{fig:deltaV} displays the evolution of $\Delta\hat{V}'_k(\hat{\rho})$ as a function of the rescaled field $\hat{\rho}$ for several RG scales. At the UV scale ($k = 1000\,\mathrm{MeV}$, red curve), the correction vanishes identically, as enforced by the network architecture through the multiplicative factor of $t$ in the output layer.

At lower RG scales ($k = 200\,\mathrm{MeV}$), the correction develops a non-monotonic field dependence with a pronounced peak around $\hat{\rho} \approx 0.04$, corresponding to the region near the effective potential minimum. In the deep infrared ($k = 20\,\mathrm{MeV}$), the correction becomes highly localized near the potential minimum, capturing the subtle finite-$N$ effects that distinguish the physical $O(4)$ theory from its large-$N$ approximation.

The right panel provides a complementary view through a heat map of $\Delta\hat{V}'(t,\hat{\rho})$ over the full training domain, where $t = \ln(k/\Lambda)$ ranges from $0$ (UV) to $-4$ (IR). The smooth color gradients demonstrate the network's ability to interpolate continuously across the $(t, \hat{\rho})$ space without artificial discontinuities or oscillations. The largest finite-$N$ corrections (red regions) are concentrated at small $\hat{\rho}$ and intermediate RG scales ($t \sim -0.5$ to $-1.5$), corresponding to momentum scales $k \sim 200$--$600\,\mathrm{MeV}$ where the sigma-mode fluctuations contribute most significantly to the RG flow.

\subsubsection{Physical observables}

\begin{figure*}[t]
	\centering
	\hspace{-1em}\includegraphics[width=0.48\textwidth]{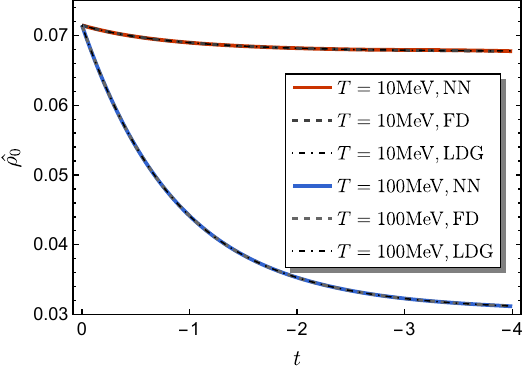}\includegraphics[width=0.48\textwidth]{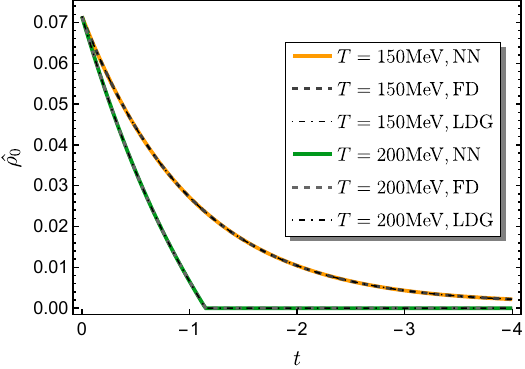}
	\caption{RG evolution of the order parameter $\hat{\rho}_0$ (position of the effective potential minimum defined by $\hat{V}'(\hat{\rho}_0) = 0$) as a function of RG time $t = \ln(k/\Lambda)$ for the $O(4)$ model. Left panel: Broken phase at $T = 10\,\mathrm{MeV}$ (red) and $T = 100\,\mathrm{MeV}$ (blue), where $\hat{\rho}_0$ flows to a finite infrared value, indicating spontaneous symmetry breaking. Right panel: Critical and symmetric phase at $T = 150\,\mathrm{MeV}$ (orange) and $T = 200\,\mathrm{MeV}$ (green), where $\hat{\rho}_0$ flows to zero, signaling restoration of $O(4)$ symmetry. Neural network results (solid lines) are compared with Finite Difference BDF (FD, dashed) and Local Discontinuous Galerkin (LDG, dot-dashed) methods.}
	\label{fig:rho0_flow}
\end{figure*}

\begin{figure*}[t]
	\centering
	\hspace{-1em}\includegraphics[width=0.48\textwidth]{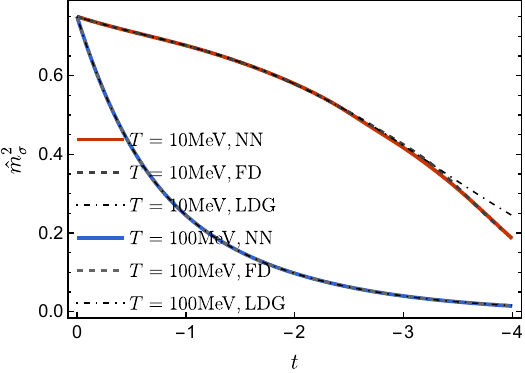}\includegraphics[width=0.48\textwidth]{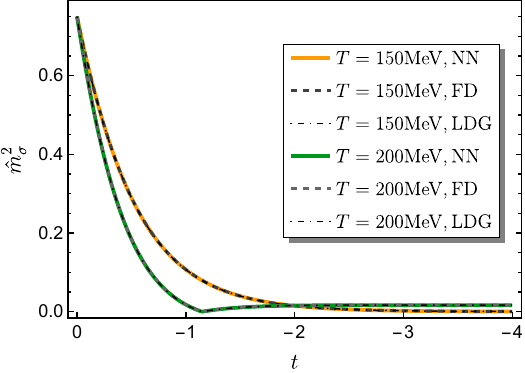}
	\caption{RG evolution of the dimensionless sigma mass squared $\hat{m}_\sigma^2 = \hat{V}'(\hat{\rho}_0) + 2\hat{\rho}_0\hat{V}''(\hat{\rho}_0)$ evaluated at the potential minimum as a function of RG time $t = \ln(k/\Lambda)$ for the $O(4)$ model. Left panel: Broken phase at $T = 10\,\mathrm{MeV}$ (red) and $T = 100\,\mathrm{MeV}$ (blue). Right panel: Critical and symmetric phase at $T = 150\,\mathrm{MeV}$ (orange) and $T = 200\,\mathrm{MeV}$ (green). Neural network results (solid lines) show excellent agreement with Finite Difference BDF (FD, dashed) and Local Discontinuous Galerkin (LDG, dot-dashed) methods.}
	\label{fig:msig2_flow}
\end{figure*}

Beyond the full field-dependent effective potential, we also extract physical observables that characterize the thermodynamic state of the system. We present the RG evolution of the order parameter $\hat{\rho}_0$ (the location of the potential minimum) and the sigma mass squared $\hat{m}_\sigma^2 = \hat{V}'(\hat{\rho}_0) + 2\hat{\rho}_0\hat{V}''(\hat{\rho}_0)$ evaluated at this minimum in \Fig{fig:rho0_flow} and \Fig{fig:msig2_flow}.

The order parameter $\hat{\rho}_0$, shown in \Fig{fig:rho0_flow}, indicates whether there is a spontaneous symmetry breaking. In the broken phase (left panel, $T = 10\,\mathrm{MeV}$ and $T = 100\,\mathrm{MeV}$), $\hat{\rho}_0$ decreases from its UV initial value ($\hat{\rho}_0 \approx 0.071$) but stays finite in the infrared. The evolution is slower at lower temperature, reflecting the overall $T$ prefactor in the flow equation \Eq{eq:flow_potential_modelA} (equivalently \Eq{eq:flow_potential_rescale} in our rescaled variables). In the critical and symmetric regime (right panel, $T = 150\,\mathrm{MeV}$ and $T = 200\,\mathrm{MeV}$), $\hat{\rho}_0$ is driven to zero, indicating restoration of the $O(4)$ symmetry. This trend is also visible directly in the field profiles such as \Fig{fig:T100_solutions} and \Fig{fig:T10_solutions}; as $k$ decreases, the flat region in $\hat{V}'_k(\hat{\rho})$ expands and pushes the effective minimum toward smaller fields.

The sigma mass squared $\hat{m}_\sigma^2$, displayed in \Fig{fig:msig2_flow}, is extracted from the learned potential via $\hat{m}_\sigma^2 = \hat{V}'(\hat{\rho}_0) + 2\hat{\rho}_0\hat{V}''(\hat{\rho}_0)$. Since this quantity depends on a second field derivative evaluated at a running, solution-dependent location $\hat{\rho}_0(k)$, it provides a stringent test of the learned solution beyond the field profiles themselves. In the broken phase (left panel), $\hat{m}_\sigma^2$ decreases monotonically as the flow approaches convexity. The emergence of a plateau $\hat{V}'_k(\hat{\rho})\simeq \mathrm{const}$ for $0\le \hat{\rho}\le \hat{\rho}_0$ implies $\hat{V}''(\hat{\rho}_0)\to 0$ and thus $\hat{m}_\sigma^2 = 2\hat{\rho}_0\hat{V}''(\hat{\rho}_0)\to 0$ in the deep infrared $t \to - \infty$. Note that $\hat{m}_\sigma^2$ is finite at a finite value of $t$ in the broken phase, while at the critical temperature it is vanishing when $t$ is smaller than some finite value, as shown by the orange line in the right panel of \Fig{fig:msig2_flow}. Above $T_c$ ($T = 200\,\mathrm{MeV}$, green line in the right panel), $\hat{m}_\sigma^2$ first decreases rapidly to zero at the scale where $\hat{\rho}_0$ vanishes in \Fig{fig:rho0_flow}, reflecting the disappearance of the broken minimum. It then increases again as the system enters the symmetric phase, where $m_\sigma^2 = V'(0) = m_\pi^2$ and the sigma and pion modes become degenerate. The excellent agreement between the neural network and the conventional solvers across all temperatures demonstrates that the neural network captures not only $\hat{V}'$ but also its curvature reliably, even in the stiff, small-field regime where convexity restoration occurs. We note that at $T=10\,\mathrm{MeV}$, a small deviation between the neural network and the LDG method appears in the deep infrared. This discrepancy arises from the extreme numerical stiffness in the deeply broken phase, where the convexity restoration generates very steep gradients around the potential minimum. In this regime, different numerical methods are sensitive to their respective discretization schemes, and the accumulated numerical errors become most pronounced at the lowest RG scales. Nevertheless, the overall agreement remains quantitatively satisfactory.

\begin{figure*}[t]
	\centering
	\hspace{-1em}\includegraphics[width=0.465\textwidth]{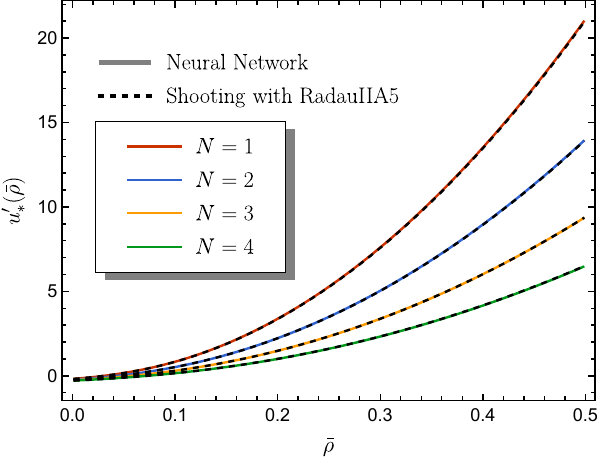}\includegraphics[width=0.48\textwidth]{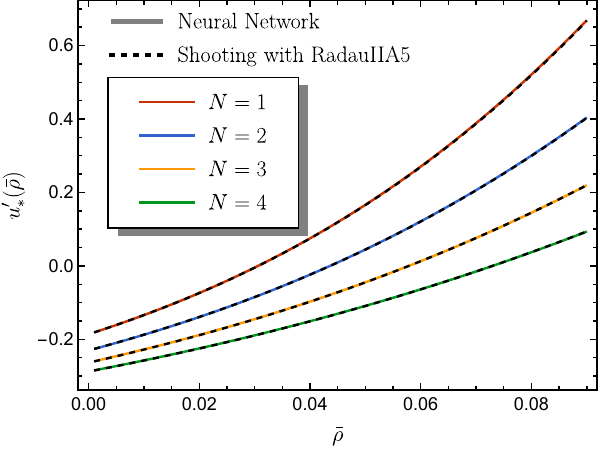}
	\caption{Fixed point solutions for $O(N)$ models with $N=1,2,3,4$ near the critical point. Left panel: Full field range showing the global structure of the fixed point potential derivatives. Right panel: Detailed view of the small field region.}
	\label{fig:fixedpoint_analysis}
\end{figure*}

\subsubsection{Wilson-Fisher fixed point}

As a further application of the neural network methodology, we solve for the Wilson-Fisher fixed point of the $O(N)$ model in $d=3$ dimensions within the LPA ($\eta=0$). The Wilson-Fisher fixed point governs the universal critical behavior of second-order phase transitions in the $O(N)$ universality class, and accurate determination of the fixed point potential is essential for computing critical exponents~\cite{Tan:2024fuq}. At the fixed point, scale invariance requires $\partial_t u(\bar{\rho})=0$, reducing the flow equation to an ordinary differential equation for the dimensionless potential derivative $u'(\bar{\rho})$. Although this problem is lower-dimensional than the full RG flow, it poses its own challenges. The fixed point ODE contains rational nonlinearities with denominators that could be zero, and numerical integration starting from small field typically encounters a singularity at finite field~\cite{Tan:2022ksv}. Conventional shooting methods require careful fine-tuning of parameters to isolate the Wilson-Fisher solution from the Gaussian fixed point or spurious multicritical solutions.

We employ the same strategy used for the finite-temperature flows. The network learns only the finite-$N$ correction $\Delta u'(\bar{\rho}) = u'(\bar{\rho}) - u'_{\mathrm{LN}}(\bar{\rho})$, where the large-$N$ fixed point solution $u'_{\mathrm{LN}}(\bar{\rho})$ is known analytically via the implicit hypergeometric expression given in Eq.~\eqref{eq:fixed_point_largeN}. This decomposition biases the network toward the physically relevant Wilson-Fisher branch and avoids collapsing onto the Gaussian fixed point. The network architecture is a simple four-layer multilayer perceptron (MLP) with 256 hidden units per layer and SiLU activations, trained by minimizing the ODE residual of the fixed point equation using finite-difference derivatives.

\Fig{fig:fixedpoint_analysis} displays the resulting fixed point potential derivative $u'_*(\bar{\rho})$ for $N=1$ (Ising), $N=2$ (XY), $N=3$ (Heisenberg), and $N=4$ (relevant for chiral symmetry restoration in two-flavor QCD). The left panel shows the global field dependence. All curves start negative at small $\bar{\rho}$ and grow monotonically, with smaller $N$ corresponding to steeper slopes. In the large-$N$ limit, the $(N-1)$ pion-mode contributions dominate the flow equation, resulting in a flatter fixed point potential profile. As $N$ decreases, the relative weight of the sigma-mode contribution increases, leading to steeper slopes in $u'_*(\bar{\rho})$.

The right panel provides a zoom-in view of the small-field region $\bar{\rho}\lesssim 0.09$, which is most sensitive to the critical properties of the phase transition. The correlation-length exponent $\nu$ can be extracted from the relevant eigenvalue of the linearized flow around the fixed point solution $u'_*(\bar{\rho})$~\cite{Tan:2022ksv}, while the anomalous dimension $\eta$ (which vanishes identically in the LPA) can be obtained by extending to higher-order truncations such as LPA$'$. The neural network results (solid gray lines) are validated against a shooting method employing the RadauIIA5 implicit Runge-Kutta integrator (dashed lines), which is well-suited for stiff boundary-value problems. See Ref.~\cite{Tan:2022ksv} for details. The agreement is excellent across all values of $N$ and the entire field range.

These results demonstrate that the neural network framework extends naturally from time-dependent RG flows to stationary fixed point equations, providing a unified machine-learning approach to fRG calculations. The ability to reliably isolate the Wilson-Fisher fixed point without manual tuning of shooting parameters or boundary conditions highlights the robustness of the neural network architecture and its potential for systematic studies of critical phenomena across different universality classes.

\subsubsection{Approximate baselines and global regularity}
\label{sec:beyond_largeN}

The fixed-point calculations above use the analytic large-$N$ solution as a reference. We isolate the role of this reference by replacing $u'_{\mathrm{LN}}$ with approximations constructed directly from the fixed-point equation. The tests are performed for the $O(1)$ model in $d=3$. The network architecture, residual loss, and optimization setup are kept unchanged, and all trained solutions are compared with the shooting solution.

In $d=3$, the fixed-point solution approaches the classical scaling form $u_*'(\bar{\rho})\to A\,\bar{\rho}^{2}$ at large field, and the amplitude $A=3\gamma$ with $\gamma\simeq 28.0608$ is known a priori from the asymptotic analysis of the fixed-point equation~\cite{Tan:2022ksv}. Although this pure scaling form is qualitatively wrong at small field, having no minimum and no zero crossing, the neural network initialized with $u'_{\rm base}=A\,\bar{\rho}^{2}$ nevertheless converges to the Wilson--Fisher solution after training. If the baseline is instead set to zero, training converges to the Gaussian fixed point. The role of the baseline is therefore to steer the optimizer away from the Gaussian solution, and no special property of the large-$N$ limit is involved.

A composite ansatz works even better. We take
\begin{align}
	u'_{\rm base}(\bar{\rho})=u'_{\rm poly}(\bar{\rho})\,[1-w(\bar{\rho})]
	+A\,\bar{\rho}^{2}\,w(\bar{\rho})\,,
	\label{eq:composite_baseline}
\end{align}
where $u'_{\rm poly}$ is the derivative of a tenth-order Taylor expansion of the fixed-point potential around $\bar{\rho}=0$ and $w(\bar{\rho})=\tfrac12\{1+\tanh[(\bar{\rho}-0.1)/0.02]\}$ is a smooth junction. The Taylor coefficients are obtained by inserting the expansion into the fixed-point equation and matching powers of $\bar{\rho}$, which reduces the equation to a set of algebraic equations for the coefficients. These equations have several roots, and we select the Wilson--Fisher one iteratively. We first solve the second-order truncation and identify its Wilson--Fisher root, then solve the third-order truncation with a root search started from the second-order coefficients, and proceed in this way order by order up to the tenth. The junction switches to the asymptotic form before the expansion loses validity ($\bar{\rho}\simeq0.1$--$0.15$).

\begin{figure}[t]
	\centering
	\includegraphics[width=\columnwidth]{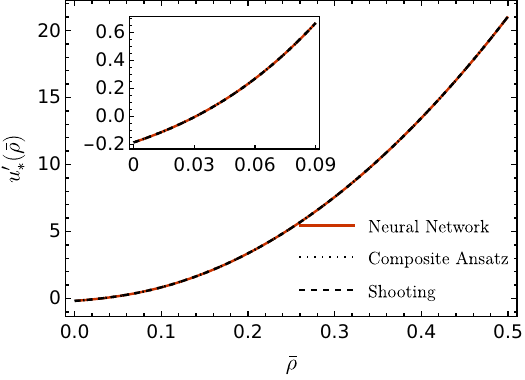}
	\caption{Fixed-point potential derivative $u_*'(\bar{\rho})$ of the $O(1)$ model in $d=3$ over the full field range, with the small-field region shown in the inset. The solution trained from the $\tanh$-composite ansatz (red solid), the ansatz itself (black dotted), and the shooting benchmark (black dashed) are indistinguishable on this scale.}
	\label{fig:baseline_composite}
\end{figure}

\begin{figure}[t]
	\centering
	\includegraphics[width=0.45\textwidth]{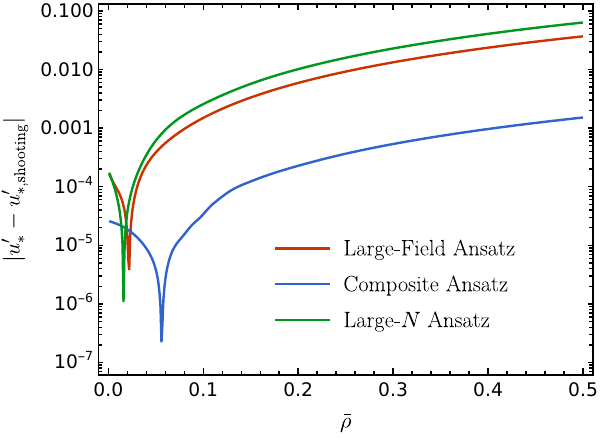}
	\caption{Absolute deviation of the trained solutions from the shooting benchmark for networks initialized with the large-field ansatz $A\bar{\rho}^{2}$ (red), the $\tanh$-composite ansatz (blue), and the large-$N$ ansatz (green).}
	\label{fig:baseline_error}
\end{figure}

The trained solution reproduces the shooting benchmark to $1.5\times10^{-3}$ over the full interval ($2.6\times10^{-5}$ at small field), an order of magnitude better than the result obtained with the large-$N$ baseline ($6.3\times10^{-2}$), as shown in Figs.~\ref{fig:baseline_composite} and~\ref{fig:baseline_error}. With a linear junction $w=\bar{\rho}/\bar{\rho}_{\max}$, training also converges to the Wilson--Fisher solution, but the polynomial part of the baseline then diverges at large field and limits the accuracy to $\sim\!1\%$. A junction that saturates before the domain boundary is therefore preferable.
\begin{figure*}[t]
	\centering
	\hspace{-1em}\includegraphics[width=0.48\textwidth]{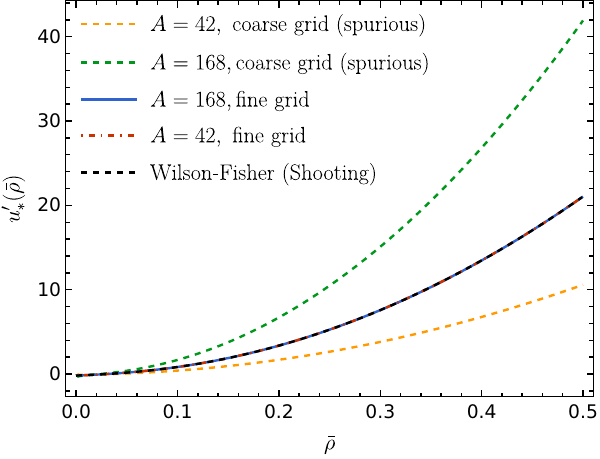}\includegraphics[width=0.48\textwidth]{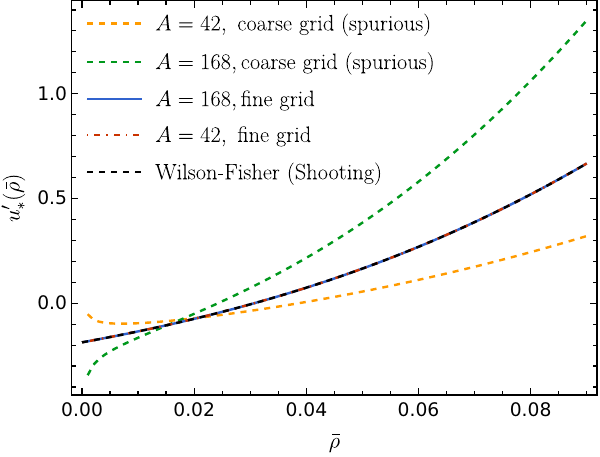}
	\caption{Recovery of the Wilson--Fisher solution from mistuned amplitudes over the full field range (left) and at small field (right). On the coarse grid, the solutions trained from $A=42$ (orange dashed) and $A=168$ (green dashed) are spurious solutions that exist only on the truncated field interval. On the refined grid ($\bar{\rho}_{\min}=10^{-4}$, $5000$ points), the solutions trained from the same two initializations (blue solid for $A=168$ and red dot-dashed for $A=42$) recover the Wilson--Fisher solution and coincide with the shooting benchmark (black dashed).}
	\label{fig:baseline_finegrid}
\end{figure*}

A single, well-characterized failure mode occurs when the pure scaling baseline $u'_{\rm base}=A\,\bar{\rho}^{2}$ is used with a strongly mistuned amplitude and without any small-field information. For example, with $A=42\simeq1.5\gamma$ or $A=168\simeq6\gamma$, a factor of two below and above the correct value $3\gamma\simeq84$, the training loss still drops to a small residual, but the trained solution is \emph{not} the Wilson--Fisher one. Its large-field behavior remains close to the input $A\,\bar{\rho}^{2}$ instead of moving to the correct amplitude. This behavior has a simple explanation. A physical fixed-point solution must be finite for all $\bar{\rho}\geq0$, and in $d=3$ this requirement leaves only two solutions, the Gaussian one $u_*'\equiv0$ and the Wilson--Fisher one. The point $\bar{\rho}=0$, however, lies outside the training interval $[\bar{\rho}_{\min},\bar{\rho}_{\max}]$. On this interval a smooth solution exists for any large-field amplitude $A$, and it can be constructed by integrating the fixed-point equation inward from $\bar{\rho}_{\max}$. All these solutions have an equally small residual loss, so training simply keeps the amplitude supplied by the baseline. Indeed, the two runs end on the solutions of amplitude $42.3$ and $167.7$, which we verified by comparison with the inward integration.

A solution with a wrong amplitude diverges toward $\bar{\rho}=0$, and the onset of this divergence can already be seen at small field in the coarse-grid curves in the right panel of Fig.~\ref{fig:baseline_finegrid}. On the coarse grid the divergence stays harmless. For $\bar{\rho}\geq10^{-3}$ the spurious solutions are still smooth and slowly varying, so their residual is small everywhere in the training domain. On the finer grid starting at $\bar{\rho}_{\min}=10^{-4}$ with $5000$ points, the same solutions vary rapidly near the lower end of the interval, their residual there becomes large, and they cease to be minima of the loss. Training in double precision with a step-decayed learning rate then recovers the Wilson--Fisher solution from both $A\simeq1.5\gamma$ and $A\simeq6\gamma$, and the trained amplitude agrees with $3\gamma$ to $0.005\%$, as shown in Fig.~\ref{fig:baseline_finegrid}. This result is also relevant to problems that do not admit special exact solutions in any limit. Even in such cases, an asymptotic analysis of the equation itself still yields the correct power-law behavior at large field. The fine-grid results show that this semi-quantitative information is already sufficient, because the physical solution is recovered even when the amplitude of the baseline is off by a factor of two.

In summary, any globally defined ansatz that keeps the optimizer away from the Gaussian solution and encodes the correct large-field asymptotics can play the same role as the large-$N$ solution. The latter is simply a convenient choice when it is available.

\section{Conclusion and Outlook}
\label{sec:conclusion}

In this work, we have developed a neural network framework for representing the effective potential in functional renormalization group calculations and demonstrated its performance for the $O(N)$ scalar field theory within the local potential approximation. By encoding the fRG flow equations directly into the loss function, the network parameters are determined so as to yield continuous solutions for the scale- and field-dependent effective potential without relying on supervised training data.

A key innovation of our approach is the decomposition of the solution into an analytically known large-$N$ contribution and a learned finite-$N$ correction. This hybrid strategy captures the dominant nonperturbative structure analytically and allows the neural network to focus on the smoother residual correction, thereby overcoming the numerical stiffness associated with convexity restoration in the broken phase. The resulting solutions agree quantitatively with established finite-difference and discontinuous Galerkin methods across all temperature regimes. We further demonstrated that the same framework applies naturally to fixed-point problems by computing the Wilson–Fisher fixed point potential in three dimensions for several values of $N$. Learning the finite-$N$ correction relative to the large-$N$ solution guides the optimization toward the physically relevant interacting fixed point and avoids convergence to the Gaussian solution, without the need for fine-tuned boundary conditions or shooting procedures. The fixed-point calculation remains accurate without an exact large-$N$ baseline. A composite baseline constructed from a local Taylor expansion and the classical large-field scaling selects the Wilson-Fisher branch and improves the accuracy relative to the large-$N$ baseline. Once the small-field region is adequately resolved, the physical solution is recovered even when the asymptotic amplitude is initially mistuned by a factor of two. This extends the framework to problems without an analytically solvable limit.

Compared with conventional grid-based numerical methods, the neural network representation offers several distinct advantages. First, it provides a continuous and differentiable parameterization of the effective potential across the entire $(t,\hat{\rho})$ domain, free from the discretization artifacts inherent in finite-difference or finite-element approaches. Second, no explicit boundary conditions need to be imposed at the boundary of a truncated computational domain in the field direction. The behavior of the effective potential at large field is not strongly constrained on physical grounds, the neural network formulation naturally accommodates this freedom by not requiring explicit boundary conditions in the field direction. Third, once trained at one temperature, the network parameters serve as an effective initialization for neighboring temperatures via transfer learning, enabling efficient systematic scans across the phase diagram. Finally, when the effective potential depends on multiple field variables, the resulting high-dimensional stiff problem renders conventional grid-based methods nearly impractical, as both memory consumption and computational cost grow combinatorially with the number of field directions. The neural network representation does not suffer from this curse of dimensionality since additional input variables can be incorporated without a fundamental change in architecture or a prohibitive increase in cost, making it a particularly promising tool for fRG applications beyond the local potential approximation.

The present results suggest several directions for future work. On the methodological side, further optimization of the network architecture and training strategy is expected to improve both efficiency and robustness. This includes more expressive input representations, such as Fourier feature embeddings~\cite{tancik2020ffn} or logarithmic scale encodings for the RG time that better capture the hierarchical structure of the flow, as well as alternative architectures beyond standard multilayer perceptrons. Improved adaptive sampling of collocation points in regions with strong gradients and the incorporation of advanced optimization techniques may further enhance performance in stiff regimes.

Beyond the present approach, other physics-driven deep learning methods offer promising alternatives for solving fRG equations. Operator-learning methods such as DeepONet~\cite{Lu_2021} and Fourier neural operators~\cite{li2021fourierneuraloperatorparametric} are designed to learn mappings between function spaces and could potentially approximate the RG flow operator that maps ultraviolet initial conditions to infrared effective actions, enabling efficient evaluation for varying microscopic parameters. In addition, the recently proposed physics-informed kernels (PIKs)~\cite{Ihssen:2024ihp,Ihssen:2025ybn,Ihssen:2025hyl,Ihssen:2026njd} approach provides a complementary perspective by encoding structural properties of renormalization group flows into reduced flow representations. Exploring possible connections between such physics-guided formulations and the operator-learning or neural-network-based approaches considered here may be an interesting direction for future studies.

From the physics perspective, the scalability of the neural network representation makes it possible to tackle fRG problems that have remained prohibitively expensive for conventional numerical methods. A particularly important direction is the extension to multi-field effective potentials, such as those arising in $2+1$ flavor QCD or in scalar theories with more complex symmetry-breaking patterns, where the effective potential depends on several independent field invariants. Equally promising is the application to the vertex expansion, where the full momentum dependence of higher-order correlation functions introduces a high-dimensional input space that has so far severely limited systematic nonperturbative studies.

Overall, this study demonstrates that neural network representations of the effective potential can serve as a robust, flexible, and scalable numerical tool for functional renormalization group calculations, complementing established methods and opening new possibilities for addressing nonperturbative problems in quantum field theory.

\begin{acknowledgments}
We thank Chuang Huang, Friederike Ihssen, Renzo Kapust and Jan M. Pawlowski, Franz R. Sattler for insightful discussions.
We thank the DEEP-IN working group at RIKEN-iTHEMS for support in the preparation of this paper.
WF is supported by the National Natural Science Foundation of China under Contract No.\ 12447102.
LW is supported by the RIKEN-TRIP initiative (RIKEN-Quantum), JSPS KAKENHI Grant No. 25H01560, and JST-BOOST Grant No. JPMJBY24H9.

\section*{Code availability}
The code accompanying this work, including the finite-temperature and Wilson-Fisher fixed-point solvers used in the present study, is publicly available in the \href{https://github.com/Yangyang-Tan/PINNforFRG}{PINNforFRG} repository~\cite{PINNforFRG_code}

\end{acknowledgments}

\appendix
\section{Derivation details for $O(N)$ flow}
\label{app:derivations}

This appendix provides detailed derivations of the flow equations and analytical solutions presented in the main text.

\subsection{Zero-temperature flow equation}
\label{app:zeroT}

Applying the Wetterich equation \eqref{eq:wetterich} to the $O(N)$ model in the LPA at zero temperature in Euclidean spacetime yields:
\begin{align}
\partial_t V_k(\rho)=\mathscr{C} k^d\left[\frac{1}{1+\bar{m}_{\sigma, k}^2}+\frac{N-1}{1+\bar{m}_{\pi, k}^2}\right]\,.
\label{eq:flow_potential_zeroT}
\end{align}
This equation serves as the basis for the fixed point analysis in \Sec{sec:fixedpoint_eq}. The dimensionless form, obtained by introducing $\bar{\rho}=k^{-(d-2)} Z_{\phi, k} \rho$ and $u(\bar{\rho})=k^{-d} V_k(\rho)$, reads:
\begin{align}
	\partial_t u(\bar{\rho})= & -d u(\bar{\rho})+(d-2+\eta) \bar{\rho} u^{\prime}(\bar{\rho}) \nonumber\\
	& +\mathscr{C}\left[\frac{1}{1+u^{\prime}+2 \bar{\rho} u^{(2)}}+\frac{N-1}{1+u^{\prime}}\right]\,.
	\label{eq:flow_potential_dimensionless}
\end{align}
Setting $\partial_t u = 0$ yields the fixed point equation \eqref{eq:fixed_point}.

\subsection{Finite temperature flow via Schwinger-Keldysh formalism}
\label{app:finite_T}

For a complete treatment of finite temperature dynamics, we employ the real-time functional renormalization group formulated on the Schwinger-Keldysh contour. The Schwinger-Keldysh formalism introduces two field components: the classical field $\phi_c$ (forward time branch) and the quantum field $\phi_q$ (difference between forward and backward branches). The scale-dependent effective action for the microscopic $O(N)$ theory~\cite{Tan:2021zid} becomes:
\begin{align}
&\Gamma_k\left[\phi_c, \phi_q\right]=\int \mathrm{d}^{d+1}x\left[Z_{\phi, k}\phi_{a,q}\left(\partial^2_t-\partial^2_i \right)\phi_{a,c}\right.\nonumber\\[2ex]
&\left.+V'_k(\rho_c)\phi_{a,q}\phi_{a,c}- q_0\epsilon\coth\left(\frac{q_0}{2T}\right)\phi_{a,q}^2 +\epsilon\phi_{a,q}\partial_t\phi_{a,c}\right]\,,
\end{align}
where $q_0$ denotes the frequency in Fourier space, $\epsilon\to 0^+$ encodes the causal $i\epsilon$ prescription, and the Keldysh noise kernel $\propto\coth(q_0/2T)$ encodes the thermal fluctuation-dissipation relation.

The corresponding flow equation incorporates thermal effects through the bosonic distribution function:
\begin{align}
\partial_t V_k(\rho)=&\mathscr{C} k^{d+1}\left[\frac{1}{2\sqrt{1+\bar{m}_{\sigma, k}^2}}\coth\left(\frac{k\sqrt{1+\bar{m}_{\sigma, k}^2}}{2T}\right)\right.\nonumber\\[2ex]
&\left.+\frac{N-1}{2\sqrt{1+\bar{m}_{\pi, k}^2}}\coth\left(\frac{k\sqrt{1+\bar{m}_{\pi, k}^2}}{2T}\right)\right]\,.
\label{eq:flow_potential_microscopic}
\end{align}

For systems with purely dissipative dynamics (Model A in the Hohenberg-Halperin classification), the effective action~\cite{Chen:2023tqc} simplifies to:
\begin{align}
	\Gamma_k\left[\phi_c, \phi_q\right]=\int dx^{d+1}&\Big[\phi_{a,q}\partial_t\phi_{a,c}-Z_{\phi, k}\phi_{a,q}\partial^2_i\phi_{a,c}\nonumber\\[2ex]
	&+V'_k(\rho_c)\phi_{a,q}\phi_{a,c}- 2T\phi_{a,q}^2 \Big]\,,
\end{align}
yielding the simplified flow equation \eqref{eq:flow_potential_modelA}. In the high temperature limit, $\coth(k\sqrt{1+\bar{m}^2}/2T)\to 2T/(k\sqrt{1+\bar{m}^2})$, and the microscopic flow \eqref{eq:flow_potential_microscopic} reduces to the Model A form.

\subsection{Method of characteristics for large-N solution}
\label{app:characteristics}

The large-$N$ flow equation \eqref{eq:flow_potential_modelA_largeN} can be solved exactly using the method of characteristics. Taking the field derivative yields:
\begin{align}
	\partial_k V'_k(\rho)=-\mathscr{C}T k^{d+1} \frac{N-1}{(k^2+V'_k(\rho))^2}V''_k(\rho)\,.
\end{align}
Defining $U_k(\rho)=V'_k(\rho)$ and rescaling $\rho\to \rho/(\mathscr{C}T(N-1))$, we obtain:
\begin{align}
	\partial_k U_k(\rho)=-k^{d+1} \frac{1}{(k^2+U_k(\rho))^2}U'_k(\rho)\,.
\end{align}

This quasi-linear first-order PDE has characteristic equations:
\begin{align}
	\frac{dk}{ds} &= 1, \quad
	\frac{d\rho}{ds} = \frac{k^{d+1}}{(k^2+U)^2}\equiv g(U,k), \quad
	\frac{dU}{ds} = 0\,.
\end{align}
The physical interpretation is clear: along each characteristic curve, $U$ remains constant while $\rho$ evolves according to $d\rho_k/dk = g(U,k)$.

Integrating from the UV scale $\Lambda$ to the running scale $k$:
\begin{align}
	\rho_k = \rho_{\Lambda} + \int_{\Lambda}^{k} g(U,k') dk'\,.
\end{align}
The integral can be evaluated analytically, yielding the exact solution \eqref{eq:largeN_solution}.

At the minimum where $U = V'_k(\rho) = 0$, the solution simplifies to:
\begin{align}
	\rho_{\text{min}}=U_\Lambda^{-1}(0)+\frac{k^{d-2}-\Lambda^{d-2}}{d-2}\,,
\end{align}
describing the RG evolution of the expectation value. For $d=3$ with linear initial conditions $U_\Lambda=\lambda(\rho-\rho_0)$: $\rho_{\text{min}}=\rho_0+k-\Lambda$.

\subsection{Fixed point ODE}
\label{app:fixedpoint_ode}

The fixed point equation \eqref{eq:fixed_point}, obtained by setting $\partial_t u = 0$ in \Eq{eq:flow_potential_dimensionless}, is a differential algebraic equation of index 1. Differentiating with respect to $\bar{\rho}$ yields a second-order ODE for $u'(\bar{\rho})$:
\begin{align}
	&\mathscr{C}\Bigg[-\frac{3 u^{(2)}+2 \bar{\rho} u^{(3)}}{\left(1+u^{\prime}+2 \bar{\rho} u^{(2)}\right)^2}-(N-1) \frac{u^{(2)}}{\left(1+u^{\prime}\right)^2}\Bigg]\nonumber\\[2ex]
	&+(-2+\eta)u'+(d-2+\eta)\bar{\rho}u^{(2)}=0\,.
\end{align}
This ODE, with appropriate boundary conditions, determines the fixed point potential and enables extraction of critical exponents characterizing the universality class.
\bibliography{ref-lib}

\begin{thebibliography}{75}%
\makeatletter
\providecommand \@ifxundefined [1]{%
 \@ifx{#1\undefined}
}%
\providecommand \@ifnum [1]{%
 \ifnum #1\expandafter \@firstoftwo
 \else \expandafter \@secondoftwo
 \fi
}%
\providecommand \@ifx [1]{%
 \ifx #1\expandafter \@firstoftwo
 \else \expandafter \@secondoftwo
 \fi
}%
\providecommand \natexlab [1]{#1}%
\providecommand \enquote  [1]{``#1''}%
\providecommand \bibnamefont  [1]{#1}%
\providecommand \bibfnamefont [1]{#1}%
\providecommand \citenamefont [1]{#1}%
\providecommand \href@noop [0]{\@secondoftwo}%
\providecommand \href [0]{\begingroup \@sanitize@url \@href}%
\providecommand \@href[1]{\@@startlink{#1}\@@href}%
\providecommand \@@href[1]{\endgroup#1\@@endlink}%
\providecommand \@sanitize@url [0]{\catcode `\\12\catcode `\$12\catcode
  `\&12\catcode `\#12\catcode `\^12\catcode `\_12\catcode `\%12\relax}%
\providecommand \@@startlink[1]{}%
\providecommand \@@endlink[0]{}%
\providecommand \url  [0]{\begingroup\@sanitize@url \@url }%
\providecommand \@url [1]{\endgroup\@href {#1}{\urlprefix }}%
\providecommand \urlprefix  [0]{URL }%
\providecommand \Eprint [0]{\href }%
\providecommand \doibase [0]{https://doi.org/}%
\providecommand \selectlanguage [0]{\@gobble}%
\providecommand \bibinfo  [0]{\@secondoftwo}%
\providecommand \bibfield  [0]{\@secondoftwo}%
\providecommand \translation [1]{[#1]}%
\providecommand \BibitemOpen [0]{}%
\providecommand \bibitemStop [0]{}%
\providecommand \bibitemNoStop [0]{.\EOS\space}%
\providecommand \EOS [0]{\spacefactor3000\relax}%
\providecommand \BibitemShut  [1]{\csname bibitem#1\endcsname}%
\let\auto@bib@innerbib\@empty
\bibitem [{\citenamefont {Wetterich}(1993)}]{Wetterich:1992yh}%
  \BibitemOpen
  \bibfield  {author} {\bibinfo {author} {\bibfnamefont {C.}~\bibnamefont
  {Wetterich}},\ }\bibfield  {title} {\bibinfo {title} {{Exact evolution
  equation for the effective potential}},\ }\href
  {https://doi.org/10.1016/0370-2693(93)90726-X} {\bibfield  {journal}
  {\bibinfo  {journal} {Phys. Lett. B}\ }\textbf {\bibinfo {volume} {301}},\
  \bibinfo {pages} {90} (\bibinfo {year} {1993})}\BibitemShut {NoStop}%
\bibitem [{\citenamefont {Wilson}(1971{\natexlab{a}})}]{Wilson:1971bg}%
  \BibitemOpen
  \bibfield  {author} {\bibinfo {author} {\bibfnamefont {K.~G.}\ \bibnamefont
  {Wilson}},\ }\bibfield  {title} {\bibinfo {title} {{Renormalization group and
  critical phenomena. 1. Renormalization group and the Kadanoff scaling
  picture}},\ }\href {https://doi.org/10.1103/PhysRevB.4.3174} {\bibfield
  {journal} {\bibinfo  {journal} {Phys. Rev. B}\ }\textbf {\bibinfo {volume}
  {4}},\ \bibinfo {pages} {3174} (\bibinfo {year}
  {1971}{\natexlab{a}})}\BibitemShut {NoStop}%
\bibitem [{\citenamefont {Wilson}(1971{\natexlab{b}})}]{Wilson:1971dh}%
  \BibitemOpen
  \bibfield  {author} {\bibinfo {author} {\bibfnamefont {K.~G.}\ \bibnamefont
  {Wilson}},\ }\bibfield  {title} {\bibinfo {title} {{Renormalization group and
  critical phenomena. 2. Phase space cell analysis of critical behavior}},\
  }\href {https://doi.org/10.1103/PhysRevB.4.3184} {\bibfield  {journal}
  {\bibinfo  {journal} {Phys. Rev. B}\ }\textbf {\bibinfo {volume} {4}},\
  \bibinfo {pages} {3184} (\bibinfo {year} {1971}{\natexlab{b}})}\BibitemShut
  {NoStop}%
\bibitem [{\citenamefont {Wilson}\ and\ \citenamefont
  {Fisher}(1972)}]{Wilson:1971dc}%
  \BibitemOpen
  \bibfield  {author} {\bibinfo {author} {\bibfnamefont {K.~G.}\ \bibnamefont
  {Wilson}}\ and\ \bibinfo {author} {\bibfnamefont {M.~E.}\ \bibnamefont
  {Fisher}},\ }\bibfield  {title} {\bibinfo {title} {{Critical exponents in
  3.99 dimensions}},\ }\href {https://doi.org/10.1103/PhysRevLett.28.240}
  {\bibfield  {journal} {\bibinfo  {journal} {Phys. Rev. Lett.}\ }\textbf
  {\bibinfo {volume} {28}},\ \bibinfo {pages} {240} (\bibinfo {year}
  {1972})}\BibitemShut {NoStop}%
\bibitem [{\citenamefont {Wilson}\ and\ \citenamefont
  {Kogut}(1974)}]{Wilson:1973jj}%
  \BibitemOpen
  \bibfield  {author} {\bibinfo {author} {\bibfnamefont {K.~G.}\ \bibnamefont
  {Wilson}}\ and\ \bibinfo {author} {\bibfnamefont {J.~B.}\ \bibnamefont
  {Kogut}},\ }\bibfield  {title} {\bibinfo {title} {{The Renormalization group
  and the epsilon expansion}},\ }\href
  {https://doi.org/10.1016/0370-1573(74)90023-4} {\bibfield  {journal}
  {\bibinfo  {journal} {Phys. Rept.}\ }\textbf {\bibinfo {volume} {12}},\
  \bibinfo {pages} {75} (\bibinfo {year} {1974})}\BibitemShut {NoStop}%
\bibitem [{\citenamefont {Berges}\ \emph {et~al.}(2002)\citenamefont {Berges},
  \citenamefont {Tetradis},\ and\ \citenamefont {Wetterich}}]{Berges:2000ew}%
  \BibitemOpen
  \bibfield  {author} {\bibinfo {author} {\bibfnamefont {J.}~\bibnamefont
  {Berges}}, \bibinfo {author} {\bibfnamefont {N.}~\bibnamefont {Tetradis}},\
  and\ \bibinfo {author} {\bibfnamefont {C.}~\bibnamefont {Wetterich}},\
  }\bibfield  {title} {\bibinfo {title} {{Nonperturbative renormalization flow
  in quantum field theory and statistical physics}},\ }\href
  {https://doi.org/10.1016/S0370-1573(01)00098-9} {\bibfield  {journal}
  {\bibinfo  {journal} {Phys. Rept.}\ }\textbf {\bibinfo {volume} {363}},\
  \bibinfo {pages} {223} (\bibinfo {year} {2002})},\ \Eprint
  {https://arxiv.org/abs/hep-ph/0005122} {arXiv:hep-ph/0005122 [hep-ph]}
  \BibitemShut {NoStop}%
\bibitem [{\citenamefont {Canet}\ and\ \citenamefont
  {Chate}(2007)}]{Canet:2006xu}%
  \BibitemOpen
  \bibfield  {author} {\bibinfo {author} {\bibfnamefont {L.}~\bibnamefont
  {Canet}}\ and\ \bibinfo {author} {\bibfnamefont {H.}~\bibnamefont {Chate}},\
  }\bibfield  {title} {\bibinfo {title} {{Non-perturbative Approach to Critical
  Dynamics}},\ }\href {https://doi.org/10.1088/1751-8113/40/9/002} {\bibfield
  {journal} {\bibinfo  {journal} {J. Phys.}\ }\textbf {\bibinfo {volume}
  {40}},\ \bibinfo {pages} {1937} (\bibinfo {year} {2007})},\ \Eprint
  {https://arxiv.org/abs/cond-mat/0610468} {arXiv:cond-mat/0610468}
  \BibitemShut {NoStop}%
\bibitem [{\citenamefont {Canet}\ \emph {et~al.}(2011)\citenamefont {Canet},
  \citenamefont {Chate},\ and\ \citenamefont {Delamotte}}]{Canet:2011wf}%
  \BibitemOpen
  \bibfield  {author} {\bibinfo {author} {\bibfnamefont {L.}~\bibnamefont
  {Canet}}, \bibinfo {author} {\bibfnamefont {H.}~\bibnamefont {Chate}},\ and\
  \bibinfo {author} {\bibfnamefont {B.}~\bibnamefont {Delamotte}},\ }\bibfield
  {title} {\bibinfo {title} {{General framework of the non-perturbative
  renormalization group for non-equilibrium steady states}},\ }\href
  {https://doi.org/10.1088/1751-8113/44/49/495001} {\bibfield  {journal}
  {\bibinfo  {journal} {J. Phys. A}\ }\textbf {\bibinfo {volume} {44}},\
  \bibinfo {pages} {495001} (\bibinfo {year} {2011})},\ \Eprint
  {https://arxiv.org/abs/1106.4129} {arXiv:1106.4129 [cond-mat.stat-mech]}
  \BibitemShut {NoStop}%
\bibitem [{\citenamefont {Mesterh\'azy}\ \emph {et~al.}(2013)\citenamefont
  {Mesterh\'azy}, \citenamefont {Stockemer}, \citenamefont {Palhares},\ and\
  \citenamefont {Berges}}]{Mesterhazy:2013naa}%
  \BibitemOpen
  \bibfield  {author} {\bibinfo {author} {\bibfnamefont {D.}~\bibnamefont
  {Mesterh\'azy}}, \bibinfo {author} {\bibfnamefont {J.~H.}\ \bibnamefont
  {Stockemer}}, \bibinfo {author} {\bibfnamefont {L.~F.}\ \bibnamefont
  {Palhares}},\ and\ \bibinfo {author} {\bibfnamefont {J.}~\bibnamefont
  {Berges}},\ }\bibfield  {title} {\bibinfo {title} {{Dynamic universality
  class of Model C from the functional renormalization group}},\ }\href
  {https://doi.org/10.1103/PhysRevB.88.174301} {\bibfield  {journal} {\bibinfo
  {journal} {Phys. Rev. B}\ }\textbf {\bibinfo {volume} {88}},\ \bibinfo
  {pages} {174301} (\bibinfo {year} {2013})},\ \Eprint
  {https://arxiv.org/abs/1307.1700} {arXiv:1307.1700 [cond-mat.stat-mech]}
  \BibitemShut {NoStop}%
\bibitem [{\citenamefont {Bluhm}\ \emph {et~al.}(2019)\citenamefont {Bluhm},
  \citenamefont {Jiang}, \citenamefont {Nahrgang}, \citenamefont {Pawlowski},
  \citenamefont {Rennecke},\ and\ \citenamefont {Wink}}]{Bluhm:2018qkf}%
  \BibitemOpen
  \bibfield  {author} {\bibinfo {author} {\bibfnamefont {M.}~\bibnamefont
  {Bluhm}}, \bibinfo {author} {\bibfnamefont {Y.}~\bibnamefont {Jiang}},
  \bibinfo {author} {\bibfnamefont {M.}~\bibnamefont {Nahrgang}}, \bibinfo
  {author} {\bibfnamefont {J.~M.}\ \bibnamefont {Pawlowski}}, \bibinfo {author}
  {\bibfnamefont {F.}~\bibnamefont {Rennecke}},\ and\ \bibinfo {author}
  {\bibfnamefont {N.}~\bibnamefont {Wink}},\ }\bibfield  {title} {\bibinfo
  {title} {{Time-evolution of fluctuations as signal of the phase transition
  dynamics in a QCD-assisted transport approach}},\ }\bibfield  {booktitle}
  {\emph {\bibinfo {booktitle} {{Proceedings, 27th International Conference on
  Ultrarelativistic Nucleus-Nucleus Collisions (Quark Matter 2018): Venice,
  Italy, May 14-19, 2018}}},\ }\href
  {https://doi.org/10.1016/j.nuclphysa.2018.09.058} {\bibfield  {journal}
  {\bibinfo  {journal} {Nucl. Phys. A}\ }\textbf {\bibinfo {volume} {982}},\
  \bibinfo {pages} {871} (\bibinfo {year} {2019})},\ \Eprint
  {https://arxiv.org/abs/1808.01377} {arXiv:1808.01377 [hep-ph]} \BibitemShut
  {NoStop}%
\bibitem [{\citenamefont {Roth}\ and\ \citenamefont {von
  Smekal}(2023)}]{Roth:2023wbp}%
  \BibitemOpen
  \bibfield  {author} {\bibinfo {author} {\bibfnamefont {J.~V.}\ \bibnamefont
  {Roth}}\ and\ \bibinfo {author} {\bibfnamefont {L.}~\bibnamefont {von
  Smekal}},\ }\bibfield  {title} {\bibinfo {title} {{Critical dynamics in a
  real-time formulation of the functional renormalization group}},\ }\href
  {https://doi.org/10.1007/JHEP10(2023)065} {\bibfield  {journal} {\bibinfo
  {journal} {JHEP}\ }\textbf {\bibinfo {volume} {10}},\ \bibinfo {pages}
  {065}},\ \Eprint {https://arxiv.org/abs/2303.11817} {arXiv:2303.11817
  [hep-ph]} \BibitemShut {NoStop}%
\bibitem [{\citenamefont {Tan}\ \emph {et~al.}(2025{\natexlab{a}})\citenamefont
  {Tan}, \citenamefont {Chen}, \citenamefont {Fu},\ and\ \citenamefont
  {Li}}]{Tan:2024fuq}%
  \BibitemOpen
  \bibfield  {author} {\bibinfo {author} {\bibfnamefont {Y.-y.}\ \bibnamefont
  {Tan}}, \bibinfo {author} {\bibfnamefont {Y.-r.}\ \bibnamefont {Chen}},
  \bibinfo {author} {\bibfnamefont {W.-j.}\ \bibnamefont {Fu}},\ and\ \bibinfo
  {author} {\bibfnamefont {W.-J.}\ \bibnamefont {Li}},\ }\bibfield  {title}
  {\bibinfo {title} {{Universality of pseudo-Goldstone damping near critical
  points}},\ }\href {https://doi.org/10.1038/s41467-025-58170-1} {\bibfield
  {journal} {\bibinfo  {journal} {Nature Commun.}\ }\textbf {\bibinfo {volume}
  {16}},\ \bibinfo {pages} {2916} (\bibinfo {year} {2025}{\natexlab{a}})},\
  \Eprint {https://arxiv.org/abs/2403.03503} {arXiv:2403.03503 [hep-th]}
  \BibitemShut {NoStop}%
\bibitem [{\citenamefont {Chen}\ \emph {et~al.}(2025)\citenamefont {Chen},
  \citenamefont {Tan},\ and\ \citenamefont {Fu}}]{Chen:2024lzz}%
  \BibitemOpen
  \bibfield  {author} {\bibinfo {author} {\bibfnamefont {Y.-r.}\ \bibnamefont
  {Chen}}, \bibinfo {author} {\bibfnamefont {Y.-y.}\ \bibnamefont {Tan}},\ and\
  \bibinfo {author} {\bibfnamefont {W.-j.}\ \bibnamefont {Fu}},\ }\bibfield
  {title} {\bibinfo {title} {{Critical dynamics of model H within the real-time
  FRG approach}},\ }\href {https://doi.org/10.1103/PhysRevD.111.094025}
  {\bibfield  {journal} {\bibinfo  {journal} {Phys. Rev. D}\ }\textbf {\bibinfo
  {volume} {111}},\ \bibinfo {pages} {094025} (\bibinfo {year} {2025})},\
  \Eprint {https://arxiv.org/abs/2406.00679} {arXiv:2406.00679 [hep-ph]}
  \BibitemShut {NoStop}%
\bibitem [{\citenamefont {Roth}\ \emph
  {et~al.}(2025{\natexlab{a}})\citenamefont {Roth}, \citenamefont {Ye},
  \citenamefont {Schlichting},\ and\ \citenamefont {von
  Smekal}}]{Roth:2024rbi}%
  \BibitemOpen
  \bibfield  {author} {\bibinfo {author} {\bibfnamefont {J.~V.}\ \bibnamefont
  {Roth}}, \bibinfo {author} {\bibfnamefont {Y.}~\bibnamefont {Ye}}, \bibinfo
  {author} {\bibfnamefont {S.}~\bibnamefont {Schlichting}},\ and\ \bibinfo
  {author} {\bibfnamefont {L.}~\bibnamefont {von Smekal}},\ }\bibfield  {title}
  {\bibinfo {title} {{Dynamic critical behavior of the chiral phase transition
  from the real-time functional renormalization group}},\ }\href
  {https://doi.org/10.1007/JHEP01(2025)118} {\bibfield  {journal} {\bibinfo
  {journal} {JHEP}\ }\textbf {\bibinfo {volume} {01}},\ \bibinfo {pages}
  {118}},\ \Eprint {https://arxiv.org/abs/2403.04573} {arXiv:2403.04573
  [hep-ph]} \BibitemShut {NoStop}%
\bibitem [{\citenamefont {Roth}\ \emph
  {et~al.}(2025{\natexlab{b}})\citenamefont {Roth}, \citenamefont {Ye},
  \citenamefont {Schlichting},\ and\ \citenamefont {von
  Smekal}}]{Roth:2024hcu}%
  \BibitemOpen
  \bibfield  {author} {\bibinfo {author} {\bibfnamefont {J.~V.}\ \bibnamefont
  {Roth}}, \bibinfo {author} {\bibfnamefont {Y.}~\bibnamefont {Ye}}, \bibinfo
  {author} {\bibfnamefont {S.}~\bibnamefont {Schlichting}},\ and\ \bibinfo
  {author} {\bibfnamefont {L.}~\bibnamefont {von Smekal}},\ }\bibfield  {title}
  {\bibinfo {title} {{Universal critical dynamics near the chiral phase
  transition and the QCD critical point}},\ }\href
  {https://doi.org/10.1103/PhysRevD.111.L111901} {\bibfield  {journal}
  {\bibinfo  {journal} {Phys. Rev. D}\ }\textbf {\bibinfo {volume} {111}},\
  \bibinfo {pages} {L111901} (\bibinfo {year} {2025}{\natexlab{b}})},\ \Eprint
  {https://arxiv.org/abs/2409.14470} {arXiv:2409.14470 [hep-ph]} \BibitemShut
  {NoStop}%
\bibitem [{\citenamefont {Braun}\ \emph {et~al.}(2016)\citenamefont {Braun},
  \citenamefont {Fister}, \citenamefont {Pawlowski},\ and\ \citenamefont
  {Rennecke}}]{Braun:2014ata}%
  \BibitemOpen
  \bibfield  {author} {\bibinfo {author} {\bibfnamefont {J.}~\bibnamefont
  {Braun}}, \bibinfo {author} {\bibfnamefont {L.}~\bibnamefont {Fister}},
  \bibinfo {author} {\bibfnamefont {J.~M.}\ \bibnamefont {Pawlowski}},\ and\
  \bibinfo {author} {\bibfnamefont {F.}~\bibnamefont {Rennecke}},\ }\bibfield
  {title} {\bibinfo {title} {{From Quarks and Gluons to Hadrons: Chiral
  Symmetry Breaking in Dynamical QCD}},\ }\href
  {https://doi.org/10.1103/PhysRevD.94.034016} {\bibfield  {journal} {\bibinfo
  {journal} {Phys. Rev. D}\ }\textbf {\bibinfo {volume} {94}},\ \bibinfo
  {pages} {034016} (\bibinfo {year} {2016})},\ \Eprint
  {https://arxiv.org/abs/1412.1045} {arXiv:1412.1045 [hep-ph]} \BibitemShut
  {NoStop}%
\bibitem [{\citenamefont {Mitter}\ \emph {et~al.}(2015)\citenamefont {Mitter},
  \citenamefont {Pawlowski},\ and\ \citenamefont
  {Strodthoff}}]{Mitter:2014wpa}%
  \BibitemOpen
  \bibfield  {author} {\bibinfo {author} {\bibfnamefont {M.}~\bibnamefont
  {Mitter}}, \bibinfo {author} {\bibfnamefont {J.~M.}\ \bibnamefont
  {Pawlowski}},\ and\ \bibinfo {author} {\bibfnamefont {N.}~\bibnamefont
  {Strodthoff}},\ }\bibfield  {title} {\bibinfo {title} {{Chiral symmetry
  breaking in continuum QCD}},\ }\href
  {https://doi.org/10.1103/PhysRevD.91.054035} {\bibfield  {journal} {\bibinfo
  {journal} {Phys. Rev. D}\ }\textbf {\bibinfo {volume} {91}},\ \bibinfo
  {pages} {054035} (\bibinfo {year} {2015})},\ \Eprint
  {https://arxiv.org/abs/1411.7978} {arXiv:1411.7978 [hep-ph]} \BibitemShut
  {NoStop}%
\bibitem [{\citenamefont {Rennecke}(2015)}]{Rennecke:2015eba}%
  \BibitemOpen
  \bibfield  {author} {\bibinfo {author} {\bibfnamefont {F.}~\bibnamefont
  {Rennecke}},\ }\bibfield  {title} {\bibinfo {title} {{Vacuum structure of
  vector mesons in QCD}},\ }\href {https://doi.org/10.1103/PhysRevD.92.076012}
  {\bibfield  {journal} {\bibinfo  {journal} {Phys. Rev. D}\ }\textbf {\bibinfo
  {volume} {92}},\ \bibinfo {pages} {076012} (\bibinfo {year} {2015})},\
  \Eprint {https://arxiv.org/abs/1504.03585} {arXiv:1504.03585 [hep-ph]}
  \BibitemShut {NoStop}%
\bibitem [{\citenamefont {Cyrol}\ \emph {et~al.}(2016)\citenamefont {Cyrol},
  \citenamefont {Fister}, \citenamefont {Mitter}, \citenamefont {Pawlowski},\
  and\ \citenamefont {Strodthoff}}]{Cyrol:2016tym}%
  \BibitemOpen
  \bibfield  {author} {\bibinfo {author} {\bibfnamefont {A.~K.}\ \bibnamefont
  {Cyrol}}, \bibinfo {author} {\bibfnamefont {L.}~\bibnamefont {Fister}},
  \bibinfo {author} {\bibfnamefont {M.}~\bibnamefont {Mitter}}, \bibinfo
  {author} {\bibfnamefont {J.~M.}\ \bibnamefont {Pawlowski}},\ and\ \bibinfo
  {author} {\bibfnamefont {N.}~\bibnamefont {Strodthoff}},\ }\bibfield  {title}
  {\bibinfo {title} {{Landau gauge Yang-Mills correlation functions}},\ }\href
  {https://doi.org/10.1103/PhysRevD.94.054005} {\bibfield  {journal} {\bibinfo
  {journal} {Phys. Rev. D}\ }\textbf {\bibinfo {volume} {94}},\ \bibinfo
  {pages} {054005} (\bibinfo {year} {2016})},\ \Eprint
  {https://arxiv.org/abs/1605.01856} {arXiv:1605.01856 [hep-ph]} \BibitemShut
  {NoStop}%
\bibitem [{\citenamefont {Cyrol}\ \emph
  {et~al.}(2018{\natexlab{a}})\citenamefont {Cyrol}, \citenamefont {Mitter},
  \citenamefont {Pawlowski},\ and\ \citenamefont {Strodthoff}}]{Cyrol:2017qkl}%
  \BibitemOpen
  \bibfield  {author} {\bibinfo {author} {\bibfnamefont {A.~K.}\ \bibnamefont
  {Cyrol}}, \bibinfo {author} {\bibfnamefont {M.}~\bibnamefont {Mitter}},
  \bibinfo {author} {\bibfnamefont {J.~M.}\ \bibnamefont {Pawlowski}},\ and\
  \bibinfo {author} {\bibfnamefont {N.}~\bibnamefont {Strodthoff}},\ }\bibfield
   {title} {\bibinfo {title} {{Nonperturbative finite-temperature Yang-Mills
  theory}},\ }\href {https://doi.org/10.1103/PhysRevD.97.054015} {\bibfield
  {journal} {\bibinfo  {journal} {Phys. Rev. D}\ }\textbf {\bibinfo {volume}
  {97}},\ \bibinfo {pages} {054015} (\bibinfo {year} {2018}{\natexlab{a}})},\
  \Eprint {https://arxiv.org/abs/1708.03482} {arXiv:1708.03482 [hep-ph]}
  \BibitemShut {NoStop}%
\bibitem [{\citenamefont {Cyrol}\ \emph
  {et~al.}(2018{\natexlab{b}})\citenamefont {Cyrol}, \citenamefont {Mitter},
  \citenamefont {Pawlowski},\ and\ \citenamefont {Strodthoff}}]{Cyrol:2017ewj}%
  \BibitemOpen
  \bibfield  {author} {\bibinfo {author} {\bibfnamefont {A.~K.}\ \bibnamefont
  {Cyrol}}, \bibinfo {author} {\bibfnamefont {M.}~\bibnamefont {Mitter}},
  \bibinfo {author} {\bibfnamefont {J.~M.}\ \bibnamefont {Pawlowski}},\ and\
  \bibinfo {author} {\bibfnamefont {N.}~\bibnamefont {Strodthoff}},\ }\bibfield
   {title} {\bibinfo {title} {{Nonperturbative quark, gluon, and meson
  correlators of unquenched QCD}},\ }\href
  {https://doi.org/10.1103/PhysRevD.97.054006} {\bibfield  {journal} {\bibinfo
  {journal} {Phys. Rev. D}\ }\textbf {\bibinfo {volume} {97}},\ \bibinfo
  {pages} {054006} (\bibinfo {year} {2018}{\natexlab{b}})},\ \Eprint
  {https://arxiv.org/abs/1706.06326} {arXiv:1706.06326 [hep-ph]} \BibitemShut
  {NoStop}%
\bibitem [{\citenamefont {Fu}\ \emph {et~al.}(2020)\citenamefont {Fu},
  \citenamefont {Pawlowski},\ and\ \citenamefont {Rennecke}}]{Fu:2019hdw}%
  \BibitemOpen
  \bibfield  {author} {\bibinfo {author} {\bibfnamefont {W.-j.}\ \bibnamefont
  {Fu}}, \bibinfo {author} {\bibfnamefont {J.~M.}\ \bibnamefont {Pawlowski}},\
  and\ \bibinfo {author} {\bibfnamefont {F.}~\bibnamefont {Rennecke}},\
  }\bibfield  {title} {\bibinfo {title} {{QCD phase structure at finite
  temperature and density}},\ }\href
  {https://doi.org/10.1103/PhysRevD.101.054032} {\bibfield  {journal} {\bibinfo
   {journal} {Phys. Rev. D}\ }\textbf {\bibinfo {volume} {101}},\ \bibinfo
  {pages} {054032} (\bibinfo {year} {2020})},\ \Eprint
  {https://arxiv.org/abs/1909.02991} {arXiv:1909.02991 [hep-ph]} \BibitemShut
  {NoStop}%
\bibitem [{\citenamefont {Braun}\ \emph {et~al.}(2020)\citenamefont {Braun},
  \citenamefont {Fu}, \citenamefont {Pawlowski}, \citenamefont {Rennecke},
  \citenamefont {Rosenbl\"uh},\ and\ \citenamefont {Yin}}]{Braun:2020ada}%
  \BibitemOpen
  \bibfield  {author} {\bibinfo {author} {\bibfnamefont {J.}~\bibnamefont
  {Braun}}, \bibinfo {author} {\bibfnamefont {W.-j.}\ \bibnamefont {Fu}},
  \bibinfo {author} {\bibfnamefont {J.~M.}\ \bibnamefont {Pawlowski}}, \bibinfo
  {author} {\bibfnamefont {F.}~\bibnamefont {Rennecke}}, \bibinfo {author}
  {\bibfnamefont {D.}~\bibnamefont {Rosenbl\"uh}},\ and\ \bibinfo {author}
  {\bibfnamefont {S.}~\bibnamefont {Yin}},\ }\bibfield  {title} {\bibinfo
  {title} {{Chiral susceptibility in ( 2+1 )-flavor QCD}},\ }\href
  {https://doi.org/10.1103/PhysRevD.102.056010} {\bibfield  {journal} {\bibinfo
   {journal} {Phys. Rev. D}\ }\textbf {\bibinfo {volume} {102}},\ \bibinfo
  {pages} {056010} (\bibinfo {year} {2020})},\ \Eprint
  {https://arxiv.org/abs/2003.13112} {arXiv:2003.13112 [hep-ph]} \BibitemShut
  {NoStop}%
\bibitem [{\citenamefont {Ihssen}\ \emph {et~al.}(2024)\citenamefont {Ihssen},
  \citenamefont {Pawlowski}, \citenamefont {Sattler},\ and\ \citenamefont
  {Wink}}]{Ihssen:2024miv}%
  \BibitemOpen
  \bibfield  {author} {\bibinfo {author} {\bibfnamefont {F.}~\bibnamefont
  {Ihssen}}, \bibinfo {author} {\bibfnamefont {J.~M.}\ \bibnamefont
  {Pawlowski}}, \bibinfo {author} {\bibfnamefont {F.~R.}\ \bibnamefont
  {Sattler}},\ and\ \bibinfo {author} {\bibfnamefont {N.}~\bibnamefont
  {Wink}},\ }\href@noop {} {\bibinfo {title} {{Towards quantitative precision
  in functional QCD I}}} (\bibinfo {year} {2024}),\ \Eprint
  {https://arxiv.org/abs/2408.08413} {arXiv:2408.08413 [hep-ph]} \BibitemShut
  {NoStop}%
\bibitem [{\citenamefont {Pawlowski}\ \emph {et~al.}(2025)\citenamefont
  {Pawlowski}, \citenamefont {Rennecke},\ and\ \citenamefont
  {Sattler}}]{Pawlowski:2025jpg}%
  \BibitemOpen
  \bibfield  {author} {\bibinfo {author} {\bibfnamefont {J.~M.}\ \bibnamefont
  {Pawlowski}}, \bibinfo {author} {\bibfnamefont {F.}~\bibnamefont
  {Rennecke}},\ and\ \bibinfo {author} {\bibfnamefont {F.~R.}\ \bibnamefont
  {Sattler}},\ }\href@noop {} {\bibinfo {title} {{Inhomogeneous instabilities
  in high-density QCD}}} (\bibinfo {year} {2025}),\ \Eprint
  {https://arxiv.org/abs/2512.20510} {arXiv:2512.20510 [hep-ph]} \BibitemShut
  {NoStop}%
\bibitem [{\citenamefont {Tan}\ \emph {et~al.}(2025{\natexlab{b}})\citenamefont
  {Tan}, \citenamefont {Yin}, \citenamefont {Chen}, \citenamefont {Huang},\
  and\ \citenamefont {Fu}}]{Tan:2025bsv}%
  \BibitemOpen
  \bibfield  {author} {\bibinfo {author} {\bibfnamefont {Y.-y.}\ \bibnamefont
  {Tan}}, \bibinfo {author} {\bibfnamefont {S.}~\bibnamefont {Yin}}, \bibinfo
  {author} {\bibfnamefont {Y.-r.}\ \bibnamefont {Chen}}, \bibinfo {author}
  {\bibfnamefont {C.}~\bibnamefont {Huang}},\ and\ \bibinfo {author}
  {\bibfnamefont {W.-j.}\ \bibnamefont {Fu}},\ }\bibfield  {title} {\bibinfo
  {title} {{Real-time evolution of critical modes in the QCD phase diagram}},\
  }\href@noop {} {\  (\bibinfo {year} {2025}{\natexlab{b}})},\ \Eprint
  {https://arxiv.org/abs/2512.03614} {arXiv:2512.03614 [hep-ph]} \BibitemShut
  {NoStop}%
\bibitem [{\citenamefont {Pawlowski}(2007)}]{Pawlowski:2005xe}%
  \BibitemOpen
  \bibfield  {author} {\bibinfo {author} {\bibfnamefont {J.~M.}\ \bibnamefont
  {Pawlowski}},\ }\bibfield  {title} {\bibinfo {title} {{Aspects of the
  functional renormalisation group}},\ }\href
  {https://doi.org/10.1016/j.aop.2007.01.007} {\bibfield  {journal} {\bibinfo
  {journal} {Annals Phys.}\ }\textbf {\bibinfo {volume} {322}},\ \bibinfo
  {pages} {2831} (\bibinfo {year} {2007})},\ \Eprint
  {https://arxiv.org/abs/hep-th/0512261} {arXiv:hep-th/0512261 [hep-th]}
  \BibitemShut {NoStop}%
\bibitem [{\citenamefont {Braun}(2012)}]{Braun:2011pp}%
  \BibitemOpen
  \bibfield  {author} {\bibinfo {author} {\bibfnamefont {J.}~\bibnamefont
  {Braun}},\ }\bibfield  {title} {\bibinfo {title} {{Fermion Interactions and
  Universal Behavior in Strongly Interacting Theories}},\ }\href
  {https://doi.org/10.1088/0954-3899/39/3/033001} {\bibfield  {journal}
  {\bibinfo  {journal} {J. Phys. G}\ }\textbf {\bibinfo {volume} {39}},\
  \bibinfo {pages} {033001} (\bibinfo {year} {2012})},\ \Eprint
  {https://arxiv.org/abs/1108.4449} {arXiv:1108.4449 [hep-ph]} \BibitemShut
  {NoStop}%
\bibitem [{\citenamefont {Dupuis}\ \emph {et~al.}(2021)\citenamefont {Dupuis},
  \citenamefont {Canet}, \citenamefont {Eichhorn}, \citenamefont {Metzner},
  \citenamefont {Pawlowski}, \citenamefont {Tissier},\ and\ \citenamefont
  {Wschebor}}]{Dupuis:2020fhh}%
  \BibitemOpen
  \bibfield  {author} {\bibinfo {author} {\bibfnamefont {N.}~\bibnamefont
  {Dupuis}}, \bibinfo {author} {\bibfnamefont {L.}~\bibnamefont {Canet}},
  \bibinfo {author} {\bibfnamefont {A.}~\bibnamefont {Eichhorn}}, \bibinfo
  {author} {\bibfnamefont {W.}~\bibnamefont {Metzner}}, \bibinfo {author}
  {\bibfnamefont {J.~M.}\ \bibnamefont {Pawlowski}}, \bibinfo {author}
  {\bibfnamefont {M.}~\bibnamefont {Tissier}},\ and\ \bibinfo {author}
  {\bibfnamefont {N.}~\bibnamefont {Wschebor}},\ }\bibfield  {title} {\bibinfo
  {title} {{The nonperturbative functional renormalization group and its
  applications}},\ }\href {https://doi.org/10.1016/j.physrep.2021.01.001}
  {\bibfield  {journal} {\bibinfo  {journal} {Phys. Rept.}\ }\textbf {\bibinfo
  {volume} {910}},\ \bibinfo {pages} {1} (\bibinfo {year} {2021})},\ \Eprint
  {https://arxiv.org/abs/2006.04853} {arXiv:2006.04853 [cond-mat.stat-mech]}
  \BibitemShut {NoStop}%
\bibitem [{\citenamefont {Fu}(2022)}]{Fu:2022gou}%
  \BibitemOpen
  \bibfield  {author} {\bibinfo {author} {\bibfnamefont {W.-j.}\ \bibnamefont
  {Fu}},\ }\bibfield  {title} {\bibinfo {title} {{QCD at finite temperature and
  density within the fRG approach: an overview}},\ }\href
  {https://doi.org/10.1088/1572-9494/ac86be} {\bibfield  {journal} {\bibinfo
  {journal} {Commun. Theor. Phys.}\ }\textbf {\bibinfo {volume} {74}},\
  \bibinfo {pages} {097304} (\bibinfo {year} {2022})},\ \Eprint
  {https://arxiv.org/abs/2205.00468} {arXiv:2205.00468 [hep-ph]} \BibitemShut
  {NoStop}%
\bibitem [{\citenamefont {Boehnlein}\ \emph {et~al.}(2022)\citenamefont
  {Boehnlein} \emph {et~al.}}]{Boehnlein:2021eym}%
  \BibitemOpen
  \bibfield  {author} {\bibinfo {author} {\bibfnamefont {A.}~\bibnamefont
  {Boehnlein}} \emph {et~al.},\ }\bibfield  {title} {\bibinfo {title}
  {{Colloquium: Machine learning in nuclear physics}},\ }\href
  {https://doi.org/10.1103/RevModPhys.94.031003} {\bibfield  {journal}
  {\bibinfo  {journal} {Rev. Mod. Phys.}\ }\textbf {\bibinfo {volume} {94}},\
  \bibinfo {pages} {031003} (\bibinfo {year} {2022})},\ \Eprint
  {https://arxiv.org/abs/2112.02309} {arXiv:2112.02309 [nucl-th]} \BibitemShut
  {NoStop}%
\bibitem [{\citenamefont {Zhou}\ \emph {et~al.}(2024)\citenamefont {Zhou},
  \citenamefont {Wang}, \citenamefont {Pang},\ and\ \citenamefont
  {Shi}}]{Zhou:2023pti}%
  \BibitemOpen
  \bibfield  {author} {\bibinfo {author} {\bibfnamefont {K.}~\bibnamefont
  {Zhou}}, \bibinfo {author} {\bibfnamefont {L.}~\bibnamefont {Wang}}, \bibinfo
  {author} {\bibfnamefont {L.-G.}\ \bibnamefont {Pang}},\ and\ \bibinfo
  {author} {\bibfnamefont {S.}~\bibnamefont {Shi}},\ }\bibfield  {title}
  {\bibinfo {title} {{Exploring QCD matter in extreme conditions with Machine
  Learning}},\ }\href {https://doi.org/10.1016/j.ppnp.2023.104084} {\bibfield
  {journal} {\bibinfo  {journal} {Prog. Part. Nucl. Phys.}\ }\textbf {\bibinfo
  {volume} {135}},\ \bibinfo {pages} {104084} (\bibinfo {year} {2024})},\
  \Eprint {https://arxiv.org/abs/2303.15136} {arXiv:2303.15136 [hep-ph]}
  \BibitemShut {NoStop}%
\bibitem [{\citenamefont {Aarts}\ \emph {et~al.}(2025)\citenamefont {Aarts},
  \citenamefont {Fukushima}, \citenamefont {Hatsuda}, \citenamefont {Ipp},
  \citenamefont {Shi}, \citenamefont {Wang},\ and\ \citenamefont
  {Zhou}}]{Aarts:2025gyp}%
  \BibitemOpen
  \bibfield  {author} {\bibinfo {author} {\bibfnamefont {G.}~\bibnamefont
  {Aarts}}, \bibinfo {author} {\bibfnamefont {K.}~\bibnamefont {Fukushima}},
  \bibinfo {author} {\bibfnamefont {T.}~\bibnamefont {Hatsuda}}, \bibinfo
  {author} {\bibfnamefont {A.}~\bibnamefont {Ipp}}, \bibinfo {author}
  {\bibfnamefont {S.}~\bibnamefont {Shi}}, \bibinfo {author} {\bibfnamefont
  {L.}~\bibnamefont {Wang}},\ and\ \bibinfo {author} {\bibfnamefont
  {K.}~\bibnamefont {Zhou}},\ }\bibfield  {title} {\bibinfo {title}
  {{Physics-driven learning for inverse problems in quantum chromodynamics}},\
  }\href {https://doi.org/10.1038/s42254-024-00798-x} {\bibfield  {journal}
  {\bibinfo  {journal} {Nature Rev. Phys.}\ }\textbf {\bibinfo {volume} {7}},\
  \bibinfo {pages} {154} (\bibinfo {year} {2025})},\ \Eprint
  {https://arxiv.org/abs/2501.05580} {arXiv:2501.05580 [hep-lat]} \BibitemShut
  {NoStop}%
\bibitem [{\citenamefont {Boyda}\ \emph {et~al.}(2022)\citenamefont {Boyda}
  \emph {et~al.}}]{Boyda:2022nmh}%
  \BibitemOpen
  \bibfield  {author} {\bibinfo {author} {\bibfnamefont {D.}~\bibnamefont
  {Boyda}} \emph {et~al.},\ }\bibfield  {title} {\bibinfo {title}
  {{Applications of Machine Learning to Lattice Quantum Field Theory}},\ }in\
  \href@noop {} {\emph {\bibinfo {booktitle} {{Snowmass 2021}}}}\ (\bibinfo
  {year} {2022})\ \Eprint {https://arxiv.org/abs/2202.05838} {arXiv:2202.05838
  [hep-lat]} \BibitemShut {NoStop}%
\bibitem [{\citenamefont {Cranmer}\ \emph {et~al.}(2023)\citenamefont
  {Cranmer}, \citenamefont {Kanwar}, \citenamefont {Racani{\`e}re},
  \citenamefont {Rezende},\ and\ \citenamefont {Shanahan}}]{Cranmer:2023xbe}%
  \BibitemOpen
  \bibfield  {author} {\bibinfo {author} {\bibfnamefont {K.}~\bibnamefont
  {Cranmer}}, \bibinfo {author} {\bibfnamefont {G.}~\bibnamefont {Kanwar}},
  \bibinfo {author} {\bibfnamefont {S.}~\bibnamefont {Racani{\`e}re}}, \bibinfo
  {author} {\bibfnamefont {D.~J.}\ \bibnamefont {Rezende}},\ and\ \bibinfo
  {author} {\bibfnamefont {P.~E.}\ \bibnamefont {Shanahan}},\ }\bibfield
  {title} {\bibinfo {title} {{Advances in machine-learning-based sampling
  motivated by lattice quantum chromodynamics}},\ }\href
  {https://doi.org/10.1038/s42254-023-00616-w} {\bibfield  {journal} {\bibinfo
  {journal} {Nature Rev. Phys.}\ }\textbf {\bibinfo {volume} {5}},\ \bibinfo
  {pages} {526} (\bibinfo {year} {2023})},\ \Eprint
  {https://arxiv.org/abs/2309.01156} {arXiv:2309.01156 [hep-lat]} \BibitemShut
  {NoStop}%
\bibitem [{\citenamefont {Wang}\ \emph {et~al.}(2024)\citenamefont {Wang},
  \citenamefont {Aarts},\ and\ \citenamefont {Zhou}}]{Wang:2023exq}%
  \BibitemOpen
  \bibfield  {author} {\bibinfo {author} {\bibfnamefont {L.}~\bibnamefont
  {Wang}}, \bibinfo {author} {\bibfnamefont {G.}~\bibnamefont {Aarts}},\ and\
  \bibinfo {author} {\bibfnamefont {K.}~\bibnamefont {Zhou}},\ }\bibfield
  {title} {\bibinfo {title} {{Diffusion models as stochastic quantization in
  lattice field theory}},\ }\href {https://doi.org/10.1007/JHEP05(2024)060}
  {\bibfield  {journal} {\bibinfo  {journal} {JHEP}\ }\textbf {\bibinfo
  {volume} {05}},\ \bibinfo {pages} {060}},\ \Eprint
  {https://arxiv.org/abs/2309.17082} {arXiv:2309.17082 [hep-lat]} \BibitemShut
  {NoStop}%
\bibitem [{\citenamefont {Wang}\ \emph {et~al.}(2023)\citenamefont {Wang},
  \citenamefont {Aarts},\ and\ \citenamefont {Zhou}}]{Wang:2023sry}%
  \BibitemOpen
  \bibfield  {author} {\bibinfo {author} {\bibfnamefont {L.}~\bibnamefont
  {Wang}}, \bibinfo {author} {\bibfnamefont {G.}~\bibnamefont {Aarts}},\ and\
  \bibinfo {author} {\bibfnamefont {K.}~\bibnamefont {Zhou}},\ }\bibfield
  {title} {\bibinfo {title} {{Generative Diffusion Models for Lattice Field
  Theory}},\ }in\ \href@noop {} {\emph {\bibinfo {booktitle} {{37th Conference
  on Neural Information Processing Systems}}}}\ (\bibinfo {year} {2023})\
  \Eprint {https://arxiv.org/abs/2311.03578} {arXiv:2311.03578 [hep-lat]}
  \BibitemShut {NoStop}%
\bibitem [{\citenamefont {Zhu}\ \emph {et~al.}(2024)\citenamefont {Zhu},
  \citenamefont {Aarts}, \citenamefont {Wang}, \citenamefont {Zhou},\ and\
  \citenamefont {Wang}}]{Zhu:2024kiu}%
  \BibitemOpen
  \bibfield  {author} {\bibinfo {author} {\bibfnamefont {Q.}~\bibnamefont
  {Zhu}}, \bibinfo {author} {\bibfnamefont {G.}~\bibnamefont {Aarts}}, \bibinfo
  {author} {\bibfnamefont {W.}~\bibnamefont {Wang}}, \bibinfo {author}
  {\bibfnamefont {K.}~\bibnamefont {Zhou}},\ and\ \bibinfo {author}
  {\bibfnamefont {L.}~\bibnamefont {Wang}},\ }\bibfield  {title} {\bibinfo
  {title} {{Diffusion models for lattice gauge field simulations}},\ }in\
  \href@noop {} {\emph {\bibinfo {booktitle} {{Machine Learning and the
  Physical Sciences: Workshop at NeurIPS 2024}}}}\ (\bibinfo {year} {2024})\
  \Eprint {https://arxiv.org/abs/2410.19602} {arXiv:2410.19602 [hep-lat]}
  \BibitemShut {NoStop}%
\bibitem [{\citenamefont {Zhu}\ \emph {et~al.}(2025)\citenamefont {Zhu},
  \citenamefont {Aarts}, \citenamefont {Wang}, \citenamefont {Zhou},\ and\
  \citenamefont {Wang}}]{Zhu:2025pmw}%
  \BibitemOpen
  \bibfield  {author} {\bibinfo {author} {\bibfnamefont {Q.}~\bibnamefont
  {Zhu}}, \bibinfo {author} {\bibfnamefont {G.}~\bibnamefont {Aarts}}, \bibinfo
  {author} {\bibfnamefont {W.}~\bibnamefont {Wang}}, \bibinfo {author}
  {\bibfnamefont {K.}~\bibnamefont {Zhou}},\ and\ \bibinfo {author}
  {\bibfnamefont {L.}~\bibnamefont {Wang}},\ }\bibfield  {title} {\bibinfo
  {title} {{Physics-Conditioned Diffusion Models for Lattice Gauge Theory}},\
  }\href@noop {} {\bibfield  {journal} {\bibinfo  {journal} {Journal of High
  Energy Physics}\ } (\bibinfo {year} {2025})},\ \Eprint
  {https://arxiv.org/abs/2502.05504} {arXiv:2502.05504 [hep-lat]} \BibitemShut
  {NoStop}%
\bibitem [{\citenamefont {Aarts}\ \emph {et~al.}(2026)\citenamefont {Aarts},
  \citenamefont {Habibi}, \citenamefont {Ipp}, \citenamefont {M{\"u}ller},
  \citenamefont {Ranner}, \citenamefont {Wang}, \citenamefont {Wang},\ and\
  \citenamefont {Zhu}}]{Aarts:2026zzr}%
  \BibitemOpen
  \bibfield  {author} {\bibinfo {author} {\bibfnamefont {G.}~\bibnamefont
  {Aarts}}, \bibinfo {author} {\bibfnamefont {D.~E.}\ \bibnamefont {Habibi}},
  \bibinfo {author} {\bibfnamefont {A.}~\bibnamefont {Ipp}}, \bibinfo {author}
  {\bibfnamefont {D.~I.}\ \bibnamefont {M{\"u}ller}}, \bibinfo {author}
  {\bibfnamefont {T.~R.}\ \bibnamefont {Ranner}}, \bibinfo {author}
  {\bibfnamefont {L.}~\bibnamefont {Wang}}, \bibinfo {author} {\bibfnamefont
  {W.}~\bibnamefont {Wang}},\ and\ \bibinfo {author} {\bibfnamefont
  {Q.}~\bibnamefont {Zhu}},\ }\href@noop {} {\bibinfo {title} {{Generalizable
  Equivariant Diffusion Models for Non-Abelian Lattice Gauge Theory}}}
  (\bibinfo {year} {2026}),\ \Eprint {https://arxiv.org/abs/2601.19552}
  {arXiv:2601.19552 [hep-lat]} \BibitemShut {NoStop}%
\bibitem [{\citenamefont {Raissi}\ \emph {et~al.}(2019)\citenamefont {Raissi},
  \citenamefont {Perdikaris},\ and\ \citenamefont
  {Karniadakis}}]{RAISSI2019686}%
  \BibitemOpen
  \bibfield  {author} {\bibinfo {author} {\bibfnamefont {M.}~\bibnamefont
  {Raissi}}, \bibinfo {author} {\bibfnamefont {P.}~\bibnamefont {Perdikaris}},\
  and\ \bibinfo {author} {\bibfnamefont {G.}~\bibnamefont {Karniadakis}},\
  }\bibfield  {title} {\bibinfo {title} {Physics-informed neural networks: A
  deep learning framework for solving forward and inverse problems involving
  nonlinear partial differential equations},\ }\href
  {https://doi.org/https://doi.org/10.1016/j.jcp.2018.10.045} {\bibfield
  {journal} {\bibinfo  {journal} {Journal of Computational Physics}\ }\textbf
  {\bibinfo {volume} {378}},\ \bibinfo {pages} {686} (\bibinfo {year}
  {2019})}\BibitemShut {NoStop}%
\bibitem [{\citenamefont {Karniadakis}\ \emph {et~al.}(2021)\citenamefont
  {Karniadakis}, \citenamefont {Kevrekidis}, \citenamefont {Lu}, \citenamefont
  {Perdikaris}, \citenamefont {Wang},\ and\ \citenamefont
  {Yang}}]{Karniadakis2021}%
  \BibitemOpen
  \bibfield  {author} {\bibinfo {author} {\bibfnamefont {G.~E.}\ \bibnamefont
  {Karniadakis}}, \bibinfo {author} {\bibfnamefont {I.~G.}\ \bibnamefont
  {Kevrekidis}}, \bibinfo {author} {\bibfnamefont {L.}~\bibnamefont {Lu}},
  \bibinfo {author} {\bibfnamefont {P.}~\bibnamefont {Perdikaris}}, \bibinfo
  {author} {\bibfnamefont {S.}~\bibnamefont {Wang}},\ and\ \bibinfo {author}
  {\bibfnamefont {L.}~\bibnamefont {Yang}},\ }\bibfield  {title} {\bibinfo
  {title} {Physics-informed machine learning},\ }\bibfield  {journal} {\bibinfo
   {journal} {Nature Reviews Physics}\ }\textbf {\bibinfo {volume} {3}},\ \href
  {https://doi.org/10.1038/s42254-021-00314-5} {10.1038/s42254-021-00314-5}
  (\bibinfo {year} {2021})\BibitemShut {NoStop}%
\bibitem [{\citenamefont {Yokota}(2024)}]{Yokota:2023czk}%
  \BibitemOpen
  \bibfield  {author} {\bibinfo {author} {\bibfnamefont {T.}~\bibnamefont
  {Yokota}},\ }\bibfield  {title} {\bibinfo {title} {{Physics-informed neural
  networks for solving functional renormalization group on a lattice}},\ }\href
  {https://doi.org/10.1103/PhysRevB.109.214205} {\bibfield  {journal} {\bibinfo
   {journal} {Phys. Rev. B}\ }\textbf {\bibinfo {volume} {109}},\ \bibinfo
  {pages} {214205} (\bibinfo {year} {2024})},\ \Eprint
  {https://arxiv.org/abs/2312.16038} {arXiv:2312.16038 [cond-mat.dis-nn]}
  \BibitemShut {NoStop}%
\bibitem [{\citenamefont {Miyagawa}\ and\ \citenamefont
  {Yokota}(2024)}]{Miyagawa:2024yrw}%
  \BibitemOpen
  \bibfield  {author} {\bibinfo {author} {\bibfnamefont {T.}~\bibnamefont
  {Miyagawa}}\ and\ \bibinfo {author} {\bibfnamefont {T.}~\bibnamefont
  {Yokota}},\ }\bibfield  {title} {\bibinfo {title} {{Physics-informed Neural
  Networks for Functional Differential Equations: Cylindrical Approximation and
  Its Convergence Guarantees}},\ }in\ \href@noop {} {\emph {\bibinfo
  {booktitle} {{38th conference on Neural Information Processing Systems}}}}\
  (\bibinfo {year} {2024})\ \Eprint {https://arxiv.org/abs/2410.18153}
  {arXiv:2410.18153 [math.NA]} \BibitemShut {NoStop}%
\bibitem [{\citenamefont {Terin}(2025)}]{Terin:2024iyy}%
  \BibitemOpen
  \bibfield  {author} {\bibinfo {author} {\bibfnamefont {R.~C.}\ \bibnamefont
  {Terin}},\ }\bibfield  {title} {\bibinfo {title} {{Physics-informed neural
  networks viewpoint for solving the Dyson-Schwinger equations of quantum
  electrodynamics}},\ }\href {https://doi.org/10.21468/SciPostPhysCore.8.3.054}
  {\bibfield  {journal} {\bibinfo  {journal} {SciPost Phys. Core}\ }\textbf
  {\bibinfo {volume} {8}},\ \bibinfo {pages} {054} (\bibinfo {year} {2025})},\
  \Eprint {https://arxiv.org/abs/2411.02177} {arXiv:2411.02177 [hep-ph]}
  \BibitemShut {NoStop}%
\bibitem [{\citenamefont {Carmo~Terin}(2025)}]{CarmoTerin:2025pvs}%
  \BibitemOpen
  \bibfield  {author} {\bibinfo {author} {\bibfnamefont {R.}~\bibnamefont
  {Carmo~Terin}},\ }\href@noop {} {\bibinfo {title} {{Spectral functions in
  Minkowski quantum electrodynamics from neural reconstruction: Benchmarking
  against dispersive Dyson--Schwinger integral equations}}} (\bibinfo {year}
  {2025}),\ \Eprint {https://arxiv.org/abs/2510.24728} {arXiv:2510.24728
  [hep-ph]} \BibitemShut {NoStop}%
\bibitem [{\citenamefont {Litim}(2000)}]{Litim:2000ci}%
  \BibitemOpen
  \bibfield  {author} {\bibinfo {author} {\bibfnamefont {D.~F.}\ \bibnamefont
  {Litim}},\ }\bibfield  {title} {\bibinfo {title} {{Optimization of the exact
  renormalization group}},\ }\href
  {https://doi.org/10.1016/S0370-2693(00)00748-6} {\bibfield  {journal}
  {\bibinfo  {journal} {Phys. Lett. B}\ }\textbf {\bibinfo {volume} {486}},\
  \bibinfo {pages} {92} (\bibinfo {year} {2000})},\ \Eprint
  {https://arxiv.org/abs/hep-th/0005245} {arXiv:hep-th/0005245 [hep-th]}
  \BibitemShut {NoStop}%
\bibitem [{\citenamefont {Litim}(2001)}]{Litim:2001up}%
  \BibitemOpen
  \bibfield  {author} {\bibinfo {author} {\bibfnamefont {D.~F.}\ \bibnamefont
  {Litim}},\ }\bibfield  {title} {\bibinfo {title} {{Optimized renormalization
  group flows}},\ }\href {https://doi.org/10.1103/PhysRevD.64.105007}
  {\bibfield  {journal} {\bibinfo  {journal} {Phys. Rev. D}\ }\textbf {\bibinfo
  {volume} {64}},\ \bibinfo {pages} {105007} (\bibinfo {year} {2001})},\
  \Eprint {https://arxiv.org/abs/hep-th/0103195} {arXiv:hep-th/0103195
  [hep-th]} \BibitemShut {NoStop}%
\bibitem [{\citenamefont {Pawlowski}\ \emph {et~al.}(2017)\citenamefont
  {Pawlowski}, \citenamefont {Scherer}, \citenamefont {Schmidt},\ and\
  \citenamefont {Wetzel}}]{Pawlowski:2015mlf}%
  \BibitemOpen
  \bibfield  {author} {\bibinfo {author} {\bibfnamefont {J.~M.}\ \bibnamefont
  {Pawlowski}}, \bibinfo {author} {\bibfnamefont {M.~M.}\ \bibnamefont
  {Scherer}}, \bibinfo {author} {\bibfnamefont {R.}~\bibnamefont {Schmidt}},\
  and\ \bibinfo {author} {\bibfnamefont {S.~J.}\ \bibnamefont {Wetzel}},\
  }\bibfield  {title} {\bibinfo {title} {{Physics and the choice of regulators
  in functional renormalisation group flows}},\ }\href
  {https://doi.org/10.1016/j.aop.2017.06.017} {\bibfield  {journal} {\bibinfo
  {journal} {Annals Phys.}\ }\textbf {\bibinfo {volume} {384}},\ \bibinfo
  {pages} {165} (\bibinfo {year} {2017})},\ \Eprint
  {https://arxiv.org/abs/1512.03598} {arXiv:1512.03598 [hep-th]} \BibitemShut
  {NoStop}%
\bibitem [{\citenamefont {Balog}\ \emph {et~al.}(2019)\citenamefont {Balog},
  \citenamefont {Chat\'e}, \citenamefont {Delamotte}, \citenamefont
  {Marohnic},\ and\ \citenamefont {Wschebor}}]{Balog:2019rrg}%
  \BibitemOpen
  \bibfield  {author} {\bibinfo {author} {\bibfnamefont {I.}~\bibnamefont
  {Balog}}, \bibinfo {author} {\bibfnamefont {H.}~\bibnamefont {Chat\'e}},
  \bibinfo {author} {\bibfnamefont {B.}~\bibnamefont {Delamotte}}, \bibinfo
  {author} {\bibfnamefont {M.}~\bibnamefont {Marohnic}},\ and\ \bibinfo
  {author} {\bibfnamefont {N.}~\bibnamefont {Wschebor}},\ }\bibfield  {title}
  {\bibinfo {title} {{Convergence of Nonperturbative Approximations to the
  Renormalization Group}},\ }\href
  {https://doi.org/10.1103/PhysRevLett.123.240604} {\bibfield  {journal}
  {\bibinfo  {journal} {Phys. Rev. Lett.}\ }\textbf {\bibinfo {volume} {123}},\
  \bibinfo {pages} {240604} (\bibinfo {year} {2019})},\ \Eprint
  {https://arxiv.org/abs/1907.01829} {arXiv:1907.01829 [cond-mat.stat-mech]}
  \BibitemShut {NoStop}%
\bibitem [{\citenamefont {De~Polsi}\ \emph {et~al.}(2020)\citenamefont
  {De~Polsi}, \citenamefont {Balog}, \citenamefont {Tissier},\ and\
  \citenamefont {Wschebor}}]{DePolsi:2020pjk}%
  \BibitemOpen
  \bibfield  {author} {\bibinfo {author} {\bibfnamefont {G.}~\bibnamefont
  {De~Polsi}}, \bibinfo {author} {\bibfnamefont {I.}~\bibnamefont {Balog}},
  \bibinfo {author} {\bibfnamefont {M.}~\bibnamefont {Tissier}},\ and\ \bibinfo
  {author} {\bibfnamefont {N.}~\bibnamefont {Wschebor}},\ }\bibfield  {title}
  {\bibinfo {title} {{Precision calculation of critical exponents in the $O(N)$
  universality classes with the nonperturbative renormalization group}},\
  }\href {https://doi.org/10.1103/PhysRevE.101.042113} {\bibfield  {journal}
  {\bibinfo  {journal} {Phys. Rev. E}\ }\textbf {\bibinfo {volume} {101}},\
  \bibinfo {pages} {042113} (\bibinfo {year} {2020})},\ \Eprint
  {https://arxiv.org/abs/2001.07525} {arXiv:2001.07525 [cond-mat.stat-mech]}
  \BibitemShut {NoStop}%
\bibitem [{\citenamefont {Litim}(2002)}]{Litim:2002cf}%
  \BibitemOpen
  \bibfield  {author} {\bibinfo {author} {\bibfnamefont {D.~F.}\ \bibnamefont
  {Litim}},\ }\bibfield  {title} {\bibinfo {title} {{Critical exponents from
  optimized renormalization group flows}},\ }\href
  {https://doi.org/10.1016/S0550-3213(02)00186-4} {\bibfield  {journal}
  {\bibinfo  {journal} {Nucl. Phys. B}\ }\textbf {\bibinfo {volume} {631}},\
  \bibinfo {pages} {128} (\bibinfo {year} {2002})},\ \Eprint
  {https://arxiv.org/abs/hep-th/0203006} {arXiv:hep-th/0203006} \BibitemShut
  {NoStop}%
\bibitem [{\citenamefont {Borchardt}\ and\ \citenamefont
  {Knorr}(2015)}]{Borchardt:2015rxa}%
  \BibitemOpen
  \bibfield  {author} {\bibinfo {author} {\bibfnamefont {J.}~\bibnamefont
  {Borchardt}}\ and\ \bibinfo {author} {\bibfnamefont {B.}~\bibnamefont
  {Knorr}},\ }\bibfield  {title} {\bibinfo {title} {{Global solutions of
  functional fixed point equations via pseudospectral methods}},\ }\href
  {https://doi.org/10.1103/PhysRevD.91.105011} {\bibfield  {journal} {\bibinfo
  {journal} {Phys. Rev. D}\ }\textbf {\bibinfo {volume} {91}},\ \bibinfo
  {pages} {105011} (\bibinfo {year} {2015})},\ \bibinfo {note} {[Erratum:
  Phys.Rev.D 93, 089904 (2016)]},\ \Eprint {https://arxiv.org/abs/1502.07511}
  {arXiv:1502.07511 [hep-th]} \BibitemShut {NoStop}%
\bibitem [{\citenamefont {Borchardt}\ and\ \citenamefont
  {Knorr}(2016)}]{Borchardt:2016pif}%
  \BibitemOpen
  \bibfield  {author} {\bibinfo {author} {\bibfnamefont {J.}~\bibnamefont
  {Borchardt}}\ and\ \bibinfo {author} {\bibfnamefont {B.}~\bibnamefont
  {Knorr}},\ }\bibfield  {title} {\bibinfo {title} {{Solving functional flow
  equations with pseudo-spectral methods}},\ }\href
  {https://doi.org/10.1103/PhysRevD.94.025027} {\bibfield  {journal} {\bibinfo
  {journal} {Phys. Rev. D}\ }\textbf {\bibinfo {volume} {94}},\ \bibinfo
  {pages} {025027} (\bibinfo {year} {2016})},\ \Eprint
  {https://arxiv.org/abs/1603.06726} {arXiv:1603.06726 [hep-th]} \BibitemShut
  {NoStop}%
\bibitem [{\citenamefont {Chen}\ \emph {et~al.}(2021)\citenamefont {Chen},
  \citenamefont {Wen},\ and\ \citenamefont {Fu}}]{Chen:2021iuo}%
  \BibitemOpen
  \bibfield  {author} {\bibinfo {author} {\bibfnamefont {Y.-r.}\ \bibnamefont
  {Chen}}, \bibinfo {author} {\bibfnamefont {R.}~\bibnamefont {Wen}},\ and\
  \bibinfo {author} {\bibfnamefont {W.-j.}\ \bibnamefont {Fu}},\ }\bibfield
  {title} {\bibinfo {title} {{Critical behaviors of the O(4) and Z(2)
  symmetries in the QCD phase diagram}},\ }\href
  {https://doi.org/10.1103/PhysRevD.104.054009} {\bibfield  {journal} {\bibinfo
   {journal} {Phys. Rev. D}\ }\textbf {\bibinfo {volume} {104}},\ \bibinfo
  {pages} {054009} (\bibinfo {year} {2021})},\ \Eprint
  {https://arxiv.org/abs/2101.08484} {arXiv:2101.08484 [hep-ph]} \BibitemShut
  {NoStop}%
\bibitem [{\citenamefont {Grossi}\ and\ \citenamefont
  {Wink}(2023)}]{Grossi:2019urj}%
  \BibitemOpen
  \bibfield  {author} {\bibinfo {author} {\bibfnamefont {E.}~\bibnamefont
  {Grossi}}\ and\ \bibinfo {author} {\bibfnamefont {N.}~\bibnamefont {Wink}},\
  }\bibfield  {title} {\bibinfo {title} {{Resolving phase transitions with
  discontinuous Galerkin methods}},\ }\href
  {https://doi.org/10.21468/SciPostPhysCore.6.4.071} {\bibfield  {journal}
  {\bibinfo  {journal} {SciPost Phys. Core}\ }\textbf {\bibinfo {volume} {6}},\
  \bibinfo {pages} {071} (\bibinfo {year} {2023})},\ \Eprint
  {https://arxiv.org/abs/1903.09503} {arXiv:1903.09503 [hep-th]} \BibitemShut
  {NoStop}%
\bibitem [{\citenamefont {Sattler}\ and\ \citenamefont
  {Pawlowski}(2024)}]{Sattler:2024ozv}%
  \BibitemOpen
  \bibfield  {author} {\bibinfo {author} {\bibfnamefont {F.~R.}\ \bibnamefont
  {Sattler}}\ and\ \bibinfo {author} {\bibfnamefont {J.~M.}\ \bibnamefont
  {Pawlowski}},\ }\href@noop {} {\bibinfo {title} {{DiFfRG: A Discretisation
  Framework for functional Renormalisation Group flows}}} (\bibinfo {year}
  {2024}),\ \Eprint {https://arxiv.org/abs/2412.13043} {arXiv:2412.13043
  [hep-ph]} \BibitemShut {NoStop}%
\bibitem [{\citenamefont {Zorbach}\ \emph {et~al.}(2024)\citenamefont
  {Zorbach}, \citenamefont {Koenigstein},\ and\ \citenamefont
  {Braun}}]{Zorbach:2024rre}%
  \BibitemOpen
  \bibfield  {author} {\bibinfo {author} {\bibfnamefont {N.}~\bibnamefont
  {Zorbach}}, \bibinfo {author} {\bibfnamefont {A.}~\bibnamefont
  {Koenigstein}},\ and\ \bibinfo {author} {\bibfnamefont {J.}~\bibnamefont
  {Braun}},\ }\href@noop {} {\bibinfo {title} {{Functional Renormalization
  Group meets Computational Fluid Dynamics: RG flows in a multi-dimensional
  field space}}} (\bibinfo {year} {2024}),\ \Eprint
  {https://arxiv.org/abs/2412.16053} {arXiv:2412.16053 [cond-mat.stat-mech]}
  \BibitemShut {NoStop}%
\bibitem [{\citenamefont {Fu}\ \emph {et~al.}(2023)\citenamefont {Fu},
  \citenamefont {Huang}, \citenamefont {Pawlowski},\ and\ \citenamefont
  {Tan}}]{Fu:2022uow}%
  \BibitemOpen
  \bibfield  {author} {\bibinfo {author} {\bibfnamefont {W.-j.}\ \bibnamefont
  {Fu}}, \bibinfo {author} {\bibfnamefont {C.}~\bibnamefont {Huang}}, \bibinfo
  {author} {\bibfnamefont {J.~M.}\ \bibnamefont {Pawlowski}},\ and\ \bibinfo
  {author} {\bibfnamefont {Y.-y.}\ \bibnamefont {Tan}},\ }\bibfield  {title}
  {\bibinfo {title} {{Four-quark scatterings in QCD I}},\ }\href
  {https://doi.org/10.21468/SciPostPhys.14.4.069} {\bibfield  {journal}
  {\bibinfo  {journal} {SciPost Phys.}\ }\textbf {\bibinfo {volume} {14}},\
  \bibinfo {pages} {069} (\bibinfo {year} {2023})},\ \Eprint
  {https://arxiv.org/abs/2209.13120} {arXiv:2209.13120 [hep-ph]} \BibitemShut
  {NoStop}%
\bibitem [{\citenamefont {Fu}\ \emph {et~al.}(2025)\citenamefont {Fu},
  \citenamefont {Huang}, \citenamefont {Pawlowski}, \citenamefont {Tan},\ and\
  \citenamefont {Zhou}}]{Fu:2025hcm}%
  \BibitemOpen
  \bibfield  {author} {\bibinfo {author} {\bibfnamefont {W.-j.}\ \bibnamefont
  {Fu}}, \bibinfo {author} {\bibfnamefont {C.}~\bibnamefont {Huang}}, \bibinfo
  {author} {\bibfnamefont {J.~M.}\ \bibnamefont {Pawlowski}}, \bibinfo {author}
  {\bibfnamefont {Y.-y.}\ \bibnamefont {Tan}},\ and\ \bibinfo {author}
  {\bibfnamefont {L.-j.}\ \bibnamefont {Zhou}},\ }\bibfield  {title} {\bibinfo
  {title} {{Four-quark scatterings in QCD III}},\ }\href
  {https://doi.org/10.1103/4sh5-w4yc} {\bibfield  {journal} {\bibinfo
  {journal} {Phys. Rev. D}\ }\textbf {\bibinfo {volume} {112}},\ \bibinfo
  {pages} {054047} (\bibinfo {year} {2025})},\ \Eprint
  {https://arxiv.org/abs/2502.14388} {arXiv:2502.14388 [hep-ph]} \BibitemShut
  {NoStop}%
\bibitem [{\citenamefont {Tan}\ \emph {et~al.}(2022)\citenamefont {Tan},
  \citenamefont {Chen},\ and\ \citenamefont {Fu}}]{Tan:2021zid}%
  \BibitemOpen
  \bibfield  {author} {\bibinfo {author} {\bibfnamefont {Y.-y.}\ \bibnamefont
  {Tan}}, \bibinfo {author} {\bibfnamefont {Y.-r.}\ \bibnamefont {Chen}},\ and\
  \bibinfo {author} {\bibfnamefont {W.-j.}\ \bibnamefont {Fu}},\ }\bibfield
  {title} {\bibinfo {title} {{Real-time dynamics of the $O(4)$ scalar theory
  within the fRG approach}},\ }\href
  {https://doi.org/10.21468/SciPostPhys.12.1.026} {\bibfield  {journal}
  {\bibinfo  {journal} {SciPost Phys.}\ }\textbf {\bibinfo {volume} {12}},\
  \bibinfo {pages} {026} (\bibinfo {year} {2022})},\ \Eprint
  {https://arxiv.org/abs/2107.06482} {arXiv:2107.06482 [hep-ph]} \BibitemShut
  {NoStop}%
\bibitem [{\citenamefont {Hohenberg}\ and\ \citenamefont
  {Halperin}(1977)}]{Hohenberg:1977ym}%
  \BibitemOpen
  \bibfield  {author} {\bibinfo {author} {\bibfnamefont {P.}~\bibnamefont
  {Hohenberg}}\ and\ \bibinfo {author} {\bibfnamefont {B.}~\bibnamefont
  {Halperin}},\ }\bibfield  {title} {\bibinfo {title} {{Theory of Dynamic
  Critical Phenomena}},\ }\href {https://doi.org/10.1103/RevModPhys.49.435}
  {\bibfield  {journal} {\bibinfo  {journal} {Rev. Mod. Phys.}\ }\textbf
  {\bibinfo {volume} {49}},\ \bibinfo {pages} {435} (\bibinfo {year}
  {1977})}\BibitemShut {NoStop}%
\bibitem [{\citenamefont {Chiu}\ \emph {et~al.}(2022)\citenamefont {Chiu},
  \citenamefont {Wong}, \citenamefont {Ooi}, \citenamefont {Dao},\ and\
  \citenamefont {Ong}}]{Chiu_2022}%
  \BibitemOpen
  \bibfield  {author} {\bibinfo {author} {\bibfnamefont {P.-H.}\ \bibnamefont
  {Chiu}}, \bibinfo {author} {\bibfnamefont {J.~C.}\ \bibnamefont {Wong}},
  \bibinfo {author} {\bibfnamefont {C.}~\bibnamefont {Ooi}}, \bibinfo {author}
  {\bibfnamefont {M.~H.}\ \bibnamefont {Dao}},\ and\ \bibinfo {author}
  {\bibfnamefont {Y.-S.}\ \bibnamefont {Ong}},\ }\bibfield  {title} {\bibinfo
  {title} {Can-pinn: A fast physics-informed neural network based on
  coupled-automatic--numerical differentiation method},\ }\href
  {https://doi.org/10.1016/j.cma.2022.114909} {\bibfield  {journal} {\bibinfo
  {journal} {Computer Methods in Applied Mechanics and Engineering}\ }\textbf
  {\bibinfo {volume} {395}},\ \bibinfo {pages} {114909} (\bibinfo {year}
  {2022})}\BibitemShut {NoStop}%
\bibitem [{\citenamefont {{PhysicsNeMo
  Contributors}}(2023)}]{PhysicsNeMo:2023}%
  \BibitemOpen
  \bibfield  {author} {\bibinfo {author} {\bibnamefont {{PhysicsNeMo
  Contributors}}},\ }\href {https://github.com/NVIDIA/physicsnemo} {\bibinfo
  {title} {{NVIDIA PhysicsNeMo: An open-source framework for physics-based deep
  learning in science and engineering}}} (\bibinfo {year} {2023}),\ \bibinfo
  {note} {software}\BibitemShut {NoStop}%
\bibitem [{\citenamefont {Falcon}\ and\ \citenamefont {{The PyTorch Lightning
  team}}(2019)}]{PyTorchLightning:2019}%
  \BibitemOpen
  \bibfield  {author} {\bibinfo {author} {\bibfnamefont {W.}~\bibnamefont
  {Falcon}}\ and\ \bibinfo {author} {\bibnamefont {{The PyTorch Lightning
  team}}},\ }\href {https://doi.org/10.5281/zenodo.3828935} {\bibinfo {title}
  {{PyTorch Lightning}}} (\bibinfo {year} {2019}),\ \bibinfo {note} {version
  1.4, software}\BibitemShut {NoStop}%
\bibitem [{\citenamefont {Tan}\ \emph {et~al.}(2024)\citenamefont {Tan},
  \citenamefont {Huang}, \citenamefont {Chen},\ and\ \citenamefont
  {Fu}}]{Tan:2022ksv}%
  \BibitemOpen
  \bibfield  {author} {\bibinfo {author} {\bibfnamefont {Y.-y.}\ \bibnamefont
  {Tan}}, \bibinfo {author} {\bibfnamefont {C.}~\bibnamefont {Huang}}, \bibinfo
  {author} {\bibfnamefont {Y.-r.}\ \bibnamefont {Chen}},\ and\ \bibinfo
  {author} {\bibfnamefont {W.-j.}\ \bibnamefont {Fu}},\ }\bibfield  {title}
  {\bibinfo {title} {{Criticality of the O(N) universality via global solutions
  to nonperturbative fixed-point equations}},\ }\href
  {https://doi.org/10.1140/epjc/s10052-024-13291-7} {\bibfield  {journal}
  {\bibinfo  {journal} {Eur. Phys. J. C}\ }\textbf {\bibinfo {volume} {84}},\
  \bibinfo {pages} {897} (\bibinfo {year} {2024})},\ \Eprint
  {https://arxiv.org/abs/2211.10249} {arXiv:2211.10249 [hep-ph]} \BibitemShut
  {NoStop}%
\bibitem [{\citenamefont {Tancik}\ \emph {et~al.}(2020)\citenamefont {Tancik},
  \citenamefont {Srinivasan}, \citenamefont {Mildenhall}, \citenamefont
  {Fridovich-Keil}, \citenamefont {Raghavan}, \citenamefont {Singhal},
  \citenamefont {Ramamoorthi}, \citenamefont {Barron},\ and\ \citenamefont
  {Ng}}]{tancik2020ffn}%
  \BibitemOpen
  \bibfield  {author} {\bibinfo {author} {\bibfnamefont {M.}~\bibnamefont
  {Tancik}}, \bibinfo {author} {\bibfnamefont {P.~P.}\ \bibnamefont
  {Srinivasan}}, \bibinfo {author} {\bibfnamefont {B.}~\bibnamefont
  {Mildenhall}}, \bibinfo {author} {\bibfnamefont {S.}~\bibnamefont
  {Fridovich-Keil}}, \bibinfo {author} {\bibfnamefont {N.}~\bibnamefont
  {Raghavan}}, \bibinfo {author} {\bibfnamefont {U.}~\bibnamefont {Singhal}},
  \bibinfo {author} {\bibfnamefont {R.}~\bibnamefont {Ramamoorthi}}, \bibinfo
  {author} {\bibfnamefont {J.~T.}\ \bibnamefont {Barron}},\ and\ \bibinfo
  {author} {\bibfnamefont {R.}~\bibnamefont {Ng}},\ }\bibfield  {title}
  {\bibinfo {title} {Fourier features let networks learn high frequency
  functions in low dimensional domains},\ }in\ \href
  {http://arxiv.org/abs/2006.10739v1} {\emph {\bibinfo {booktitle} {Advances in
  Neural Information Processing Systems (NeurIPS)}}}\ (\bibinfo  {publisher}
  {Curran Associates, Inc.},\ \bibinfo {year} {2020})\BibitemShut {NoStop}%
\bibitem [{\citenamefont {Lu}\ \emph {et~al.}(2021)\citenamefont {Lu},
  \citenamefont {Jin}, \citenamefont {Pang}, \citenamefont {Zhang},\ and\
  \citenamefont {Karniadakis}}]{Lu_2021}%
  \BibitemOpen
  \bibfield  {author} {\bibinfo {author} {\bibfnamefont {L.}~\bibnamefont
  {Lu}}, \bibinfo {author} {\bibfnamefont {P.}~\bibnamefont {Jin}}, \bibinfo
  {author} {\bibfnamefont {G.}~\bibnamefont {Pang}}, \bibinfo {author}
  {\bibfnamefont {Z.}~\bibnamefont {Zhang}},\ and\ \bibinfo {author}
  {\bibfnamefont {G.~E.}\ \bibnamefont {Karniadakis}},\ }\bibfield  {title}
  {\bibinfo {title} {Learning nonlinear operators via deeponet based on the
  universal approximation theorem of operators},\ }\href
  {https://doi.org/10.1038/s42256-021-00302-5} {\bibfield  {journal} {\bibinfo
  {journal} {Nature Machine Intelligence}\ }\textbf {\bibinfo {volume} {3}},\
  \bibinfo {pages} {218} (\bibinfo {year} {2021})}\BibitemShut {NoStop}%
\bibitem [{\citenamefont {Li}\ \emph {et~al.}(2021)\citenamefont {Li},
  \citenamefont {Kovachki}, \citenamefont {Azizzadenesheli}, \citenamefont
  {Liu}, \citenamefont {Bhattacharya}, \citenamefont {Stuart},\ and\
  \citenamefont {Anandkumar}}]{li2021fourierneuraloperatorparametric}%
  \BibitemOpen
  \bibfield  {author} {\bibinfo {author} {\bibfnamefont {Z.}~\bibnamefont
  {Li}}, \bibinfo {author} {\bibfnamefont {N.}~\bibnamefont {Kovachki}},
  \bibinfo {author} {\bibfnamefont {K.}~\bibnamefont {Azizzadenesheli}},
  \bibinfo {author} {\bibfnamefont {B.}~\bibnamefont {Liu}}, \bibinfo {author}
  {\bibfnamefont {K.}~\bibnamefont {Bhattacharya}}, \bibinfo {author}
  {\bibfnamefont {A.}~\bibnamefont {Stuart}},\ and\ \bibinfo {author}
  {\bibfnamefont {A.}~\bibnamefont {Anandkumar}},\ }\href
  {https://arxiv.org/abs/2010.08895} {\bibinfo {title} {Fourier neural operator
  for parametric partial differential equations}} (\bibinfo {year} {2021}),\
  \Eprint {https://arxiv.org/abs/2010.08895} {arXiv:2010.08895 [cs.LG]}
  \BibitemShut {NoStop}%
\bibitem [{\citenamefont {Ihssen}\ and\ \citenamefont
  {Pawlowski}(2025{\natexlab{a}})}]{Ihssen:2024ihp}%
  \BibitemOpen
  \bibfield  {author} {\bibinfo {author} {\bibfnamefont {F.}~\bibnamefont
  {Ihssen}}\ and\ \bibinfo {author} {\bibfnamefont {J.~M.}\ \bibnamefont
  {Pawlowski}},\ }\bibfield  {title} {\bibinfo {title} {{Physics-informed
  renormalisation group flows}},\ }\href
  {https://doi.org/10.1016/j.aop.2025.170177} {\bibfield  {journal} {\bibinfo
  {journal} {Annals Phys.}\ }\textbf {\bibinfo {volume} {481}},\ \bibinfo
  {pages} {170177} (\bibinfo {year} {2025}{\natexlab{a}})},\ \Eprint
  {https://arxiv.org/abs/2409.13679} {arXiv:2409.13679 [hep-th]} \BibitemShut
  {NoStop}%
\bibitem [{\citenamefont {Ihssen}\ \emph {et~al.}(2025)\citenamefont {Ihssen},
  \citenamefont {Kapust},\ and\ \citenamefont {Pawlowski}}]{Ihssen:2025ybn}%
  \BibitemOpen
  \bibfield  {author} {\bibinfo {author} {\bibfnamefont {F.}~\bibnamefont
  {Ihssen}}, \bibinfo {author} {\bibfnamefont {R.}~\bibnamefont {Kapust}},\
  and\ \bibinfo {author} {\bibfnamefont {J.~M.}\ \bibnamefont {Pawlowski}},\
  }\href@noop {} {\bibinfo {title} {{Generative sampling with physics-informed
  kernels}}} (\bibinfo {year} {2025}),\ \Eprint
  {https://arxiv.org/abs/2510.26678} {arXiv:2510.26678 [hep-lat]} \BibitemShut
  {NoStop}%
\bibitem [{\citenamefont {Ihssen}\ and\ \citenamefont
  {Pawlowski}(2025{\natexlab{b}})}]{Ihssen:2025hyl}%
  \BibitemOpen
  \bibfield  {author} {\bibinfo {author} {\bibfnamefont {F.}~\bibnamefont
  {Ihssen}}\ and\ \bibinfo {author} {\bibfnamefont {J.~M.}\ \bibnamefont
  {Pawlowski}},\ }\href@noop {} {\bibinfo {title} {{Physics-informed operator
  flows and observables}}} (\bibinfo {year} {2025}{\natexlab{b}}),\ \Eprint
  {https://arxiv.org/abs/2507.13011} {arXiv:2507.13011 [hep-th]} \BibitemShut
  {NoStop}%
\bibitem [{\citenamefont {Ihssen}\ \emph {et~al.}(2026)\citenamefont {Ihssen},
  \citenamefont {Kapust},\ and\ \citenamefont {Pawlowski}}]{Ihssen:2026njd}%
  \BibitemOpen
  \bibfield  {author} {\bibinfo {author} {\bibfnamefont {F.}~\bibnamefont
  {Ihssen}}, \bibinfo {author} {\bibfnamefont {R.}~\bibnamefont {Kapust}},\
  and\ \bibinfo {author} {\bibfnamefont {J.~M.}\ \bibnamefont {Pawlowski}},\
  }\href@noop {} {\bibinfo {title} {{Solving sign problems with
  physics-informed kernels}}} (\bibinfo {year} {2026}),\ \Eprint
  {https://arxiv.org/abs/2603.03159} {arXiv:2603.03159 [hep-lat]} \BibitemShut
  {NoStop}%
\bibitem [{\citenamefont {Tan}\ \emph {et~al.}(2026)\citenamefont {Tan},
  \citenamefont {Fu}, \citenamefont {He},\ and\ \citenamefont
  {Wang}}]{PINNforFRG_code}%
  \BibitemOpen
  \bibfield  {author} {\bibinfo {author} {\bibfnamefont {Y.-y.}\ \bibnamefont
  {Tan}}, \bibinfo {author} {\bibfnamefont {W.-j.}\ \bibnamefont {Fu}},
  \bibinfo {author} {\bibfnamefont {L.}~\bibnamefont {He}},\ and\ \bibinfo
  {author} {\bibfnamefont {L.}~\bibnamefont {Wang}},\ }\href
  {https://github.com/Yangyang-Tan/PINNforFRG} {\bibinfo {title}
  {{PINNforFRG}}} (\bibinfo {year} {2026}),\ \bibinfo {note} {gitHub
  repository}\BibitemShut {NoStop}%
\bibitem [{\citenamefont {Chen}\ \emph {et~al.}(2024)\citenamefont {Chen},
  \citenamefont {Tan},\ and\ \citenamefont {Fu}}]{Chen:2023tqc}%
  \BibitemOpen
  \bibfield  {author} {\bibinfo {author} {\bibfnamefont {Y.-r.}\ \bibnamefont
  {Chen}}, \bibinfo {author} {\bibfnamefont {Y.-y.}\ \bibnamefont {Tan}},\ and\
  \bibinfo {author} {\bibfnamefont {W.-j.}\ \bibnamefont {Fu}},\ }\bibfield
  {title} {\bibinfo {title} {{Critical dynamics within the real-time FRG
  approach}},\ }\href {https://doi.org/10.1103/PhysRevD.109.094044} {\bibfield
  {journal} {\bibinfo  {journal} {Phys. Rev. D}\ }\textbf {\bibinfo {volume}
  {109}},\ \bibinfo {pages} {094044} (\bibinfo {year} {2024})},\ \Eprint
  {https://arxiv.org/abs/2312.05870} {arXiv:2312.05870 [hep-ph]} \BibitemShut
  {NoStop}%
\end{thebibliography}%


\end{document}